# Exciton–phonon coupling strength in single-layer MoSe$_2$ at room temperature

*Donghai Li[1], Chiara Trovatello[2], Stefano Dal Conte[2], Matthias Nuß[1], Giancarlo Soavi[3,4], Gang Wang[3], Andrea C. Ferrari[3*], Giulio Cerullo[2,5*] and Tobias Brixner[1,6*]*

[1]Institut für Physikalische und Theoretische Chemie, Universität Würzburg, Am Hubland, 97074 Würzburg, Germany

[2]Dipartimento di Fisica, Politecnico di Milano, Piazza L. da Vinci 32, I-20133 Milano, Italy

[3]Cambridge Graphene Centre, University of Cambridge, 9 JJ Thomson Avenue, Cambridge CB3 0FA, UK

[4]Institute for Solid State Physics, Abbe Center of Photonics, Friedrich-Schiller-University Jena, Max-Wien-Platz 1, 07743, Jena, Germany

[5]IFN-CNR, Piazza L. da Vinci 32, I-20133 Milano, Italy

[6]Center for Nanosystems Chemistry (CNC), Universität Würzburg, Theodor-Boveri-Weg, 97074 Würzburg, Germany

*e-mails: acf26@eng.cam.ac.uk, giulio.cerullo@polimi.it, brixner@uni-wuerzburg.de

## Abstract

**Single-layer transition metal dichalcogenides are at the center of an ever increasing research effort both in terms of fundamental physics and applications. Exciton–phonon coupling plays a key role in determining the (opto)electronic properties of these materials. However, the exciton–phonon coupling strength has not been measured at room temperature. Here, we develop two-dimensional micro-spectroscopy to determine**

**exciton–phonon coupling of single-layer MoSe$_2$. We detect beating signals as a function of waiting time *T*, induced by the coupling between the A exciton and the *A'*$_1$ optical phonon. Analysis of two-dimensional beating maps combined with simulations provides the exciton–phonon coupling. The Huang–Rhys factor of ~1 is larger than in most other inorganic semiconductor nanostructures. Our technique offers a unique tool to measure exciton–phonon coupling also in other heterogeneous semiconducting systems with a spatial resolution ~260 nm, and will provide design-relevant parameters for the development of optoelectronic devices.**

## Introduction

Layered materials (LMs)[1–4], such as single-layer (1L) transition metal dichalcogenides (1L-TMDs)[5–8], are a promising platform for new photonic and optoelectronic devices. Bulk semiconducting TMDs consist of covalently bound layers of the type MX$_2$, where M is a metal (e.g., Mo, W) and X is a chalcogen atom (e.g., S, Se), held together by van der Waals interactions[3]. When they are exfoliated or grown as 1L, quantum confinement induces an indirect-to-direct bandgap transition[5,6]. The reduced dimensionality is also responsible for high exciton binding energies (hundreds of meV)[7,8], making 1L-TMDs excellent candidates for optoelectronic devices at room temperature (RT)[2].

Exciton–phonon coupling (EXPC) plays a key role in determining the T-dependent optoelectronic and transport properties of 1L-TMDs[9–11]. It is responsible for, e.g., non-radiative exciton decay[9,10,12], limiting the fluorescence quantum yield[13], the formation of dark-exciton phonon replicas[14], and it mediates spin-flip processes, thus decreasing the lifetime of spin/valley-polarized charge carriers[15]. For T<100 K, the interaction between excitons and acoustic phonons induces linewidth broadening and dominates the excitonic resonance of 1L-TMDs[9,16,17]. The situation is different for higher T. Ref. 18 suggested that the coupling between excitons and optical phonons induces sidebands in the absorption spectrum of 1L-MoSe$_2$ at RT.

Yet the spectral signature of this coupling is obscured by inhomogeneous broadening[18]. The presence of EXPC was inferred from resonance Raman scattering[19,20] as well as time-resolved transmission measurements[21,22], where the $A'_1$ optical phonon mode was observed to couple with the A excitonic resonance. While the energetic position of excitons can be obtained from photoluminescence and the energetic position of (ground-state) phonons from Raman measurements, this does not fully characterize the system. For obtaining the complete Hamiltonian, one also requires the displacement along the phonon coordinate of the exciton-state potential energy minimum versus the ground state. This displacement is the EXPC strength, as further detailed below, that determines how strongly phonons will be excited upon an optical transition to the exciton state. To the best of our knowledge, the EXPC strength has never been determined for any 1L-TMD at RT, because overtone bands of the optical phonon mode were not detectable[19–22]. We determine the missing quantity in the present work.

Optical four-wave-mixing experiments in semiconductors provide access to coherent dynamics of excitons[23–25,10]. In photon echo experiments the polarization state of incident photons (circular or linear) allows one to uncover different mechanisms behind the signal formation[26,27]. Different level schemes can be distinguished by the polarization dependence[27–29]. Two-dimensional electronic spectroscopy (2DES) is a powerful tool to analyze light-induced coherences in molecular systems[30–33] and semiconductors[34,35]. It is a generalized version of transient absorption spectroscopy, providing frequency resolution not only for the probe step, but also for the pump[36–40]. Coherent broadband excitation of several quantum energy levels leads to wave packets that may be detected as oscillations of specific peaks in the 2D maps as a function of waiting time $T$[31,41]. Analysis of frequency, decay time, and the position of such oscillations allows one to explore the underlying energy structure and the coupling mechanism leading to level splittings[31,41,42]. In particular, Ref. 42 has theoretically proposed that an additional Fourier transform along $T$ and cutting the resulting 3D spectrum at certain beating frequencies could lead to 2D maps that are sensitive to EXPC strength.

It is challenging to apply 2DES on micro-scale samples or heterogeneous materials with localized structural domains on a µm lateral scale, because the standard phase-matching geometry requires the exciting beams to be non-collinear with respect to each other[38]. This cannot be realized simultaneously when focusing with a high-numerical-aperture (high-NA) objective, in which all incident light arrives from the same solid angle at the sample. As a result, if one chooses to employ phase matching, this necessarily requires longer focal lengths, leading to larger spot sizes and unwanted averaging over different spatial regions or crystal orientations[43]. Instead, one can also select the signal by phase cycling[44–46], which relies on detecting population-based signals as a function of inter-pulse phase combinations[44,46,47]. The collinear geometry accessible with phase cycling enables 2D micro-spectroscopy, i.e., the combination of 2DES with fluorescence microscopy, to gain additional spatial resolution[43,48].

Here, we develop 2D micro-spectroscopy to resolve the spectral features of the phonon sidebands in 1L-MoSe$_2$ at RT and determine the EXPC. We observe oscillations in 2D maps that arise from the coupling between the $A'_1$ optical phonon mode and the A exciton. From comparison with simulated 2D beating maps, we deduce a Huang–Rhys factor, $S\sim1$. This implies a large EXPC strength for 1L-MoSe$_2$, when compared with other inorganic semiconductor nanostructures, such as CdSe quantum dots[49] and rods[50], ZnSe quantum dots[51], single-wall carbon nanotubes[52], etc., most of which fall in the range of 0–0.5 [53], providing design-relevant information for the development of photonic devices based on 1L-MoSe$_2$. Our method can be extended to other 1L-TMDs and materials and, additionally, also to other important semiconducting systems, for which the ~260-nm spatial resolution of the 2D micro-spectroscopy is required, e.g., single-wall carbon nanotubes, van-der-Waals heterostructures of layered materials, layered perovskites, bulk heterojunctions, or microcavities with embedded semiconductors.

**Results and discussion**

The experimental setup is sketched in Fig. 1a. A Ti:sapphire oscillator emits 12-fs pulses at 80 MHz repetition rate. A pulse shaper generates a collinear four-pulse sequence, focused by a high-NA objective (NA = 1.4), so that a spatial resolution~260 nm is achieved. To image the sample, the laser focus is mapped by a piezo scanning stage, and the photoluminescence (PL) signal is detected by an avalanche photodiode (APD). For the 2D map, the PL intensity is detected while scanning a first coherence time $\tau$ (delay between the first two pulses), a waiting time $T$ (delay between the second and the third pulses), and a second coherence time $t$ (delay between the third and the fourth pulses, Fig. 1a). Fourier transformation over $\tau$ and $t$ results in a 2D map for every $T$ (see Methods for data acquisition details). Nonlinear signals are obtained by systematically scanning through a number of discrete phase steps for each pulse and for each pulse-delay combination, and rephasing and nonrephasing signals are retrieved as linear superpositions of differently phase-modulated data[46].

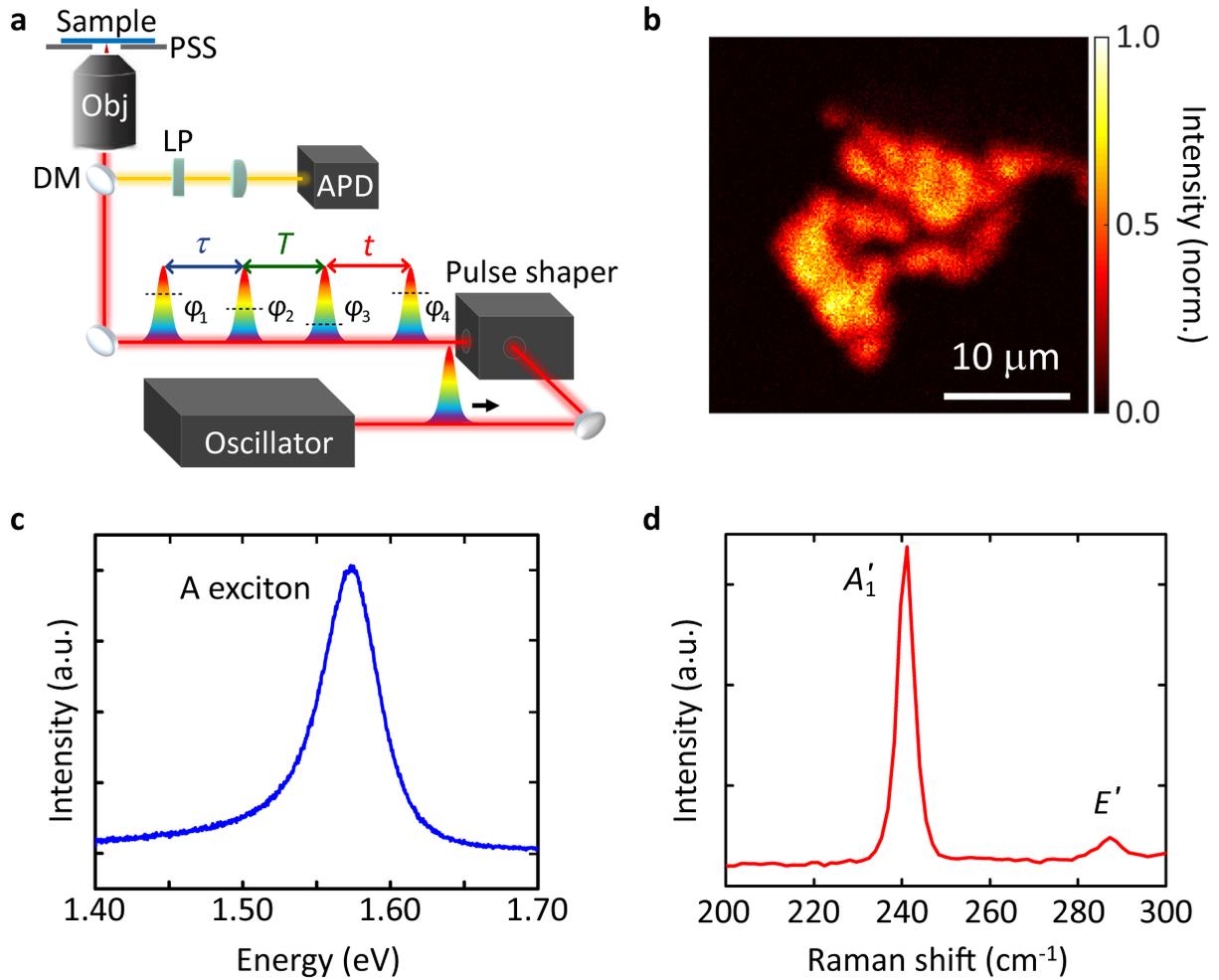

**Figure 1.** Overview of setup and the sample. **a,** Fluorescence-detected 2D micro-spectroscopy setup. Four collinear laser pulses are generated by a pulse shaper with controllable inter-pulse time delays ($\tau$, $T$, $t$) and phases ($\varphi_i$, $i$ = 1, 2, 3, 4) and focused by a high-NA objective (Obj). The position of the sample is controlled by a piezo scanning stage (PSS). The dichroic mirror (DM) under the objective is adopted to transmit the excitation beam (red) and reflect the PL signal (yellow). A long-pass filter (LP) is used to block the remaining excitation beam. The PL signal is detected by an avalanche photodiode (APD). **b,** PL map obtained with the setup of panel **a**. **c,** PL and **d,** Raman spectrum for 514 nm excitation. The peak observed in the PL spectrum corresponds to the A exciton. The Raman spectrum shows the out-of-plane $A'_1$ mode ~241 cm$^{-1}$, and the in-plane $E'$ mode ~288 cm$^{-1}$.

We investigate mechanically exfoliated 1L-MoSe$_2$ on a 200 μm fused silica substrate (see Methods for details). Figure 1b is a PL map, taken with the setup of Fig. 1a, for a representative sample. 1L-MoSe$_2$ has a direct bandgap at the K point of the Brillouin zone leading to two excitonic transitions A and B ~1.57 and 1.75 eV[54]. The PL spectrum (Fig. 1c) shows a single peak ~1.57 eV, due to the radiative recombination of A excitons[55]. The signal of the trion is much weaker than that of the neutral exciton at room temperature[25,56]. In our experiment we detect predominantly the neutral exciton. This is confirmed by the linear PL spectrum of our sample (Fig. 1c), in which the main peak is located at a position that agrees with that found for neutral excitons[55]. The Raman spectrum measured at 514 nm (Fig. 1d) shows the out-of-plane $A'_1$ mode ~241 cm$^{-1}$ with full width at half maximum (FWHM) ~4 cm$^{-1}$, and the in-plane $E'$ mode ~288 cm$^{-1}$ (FWHM ~6 cm$^{-1}$). Both PL and Raman spectra confirm that the sample is 1L-MoSe$_2$[55,19].

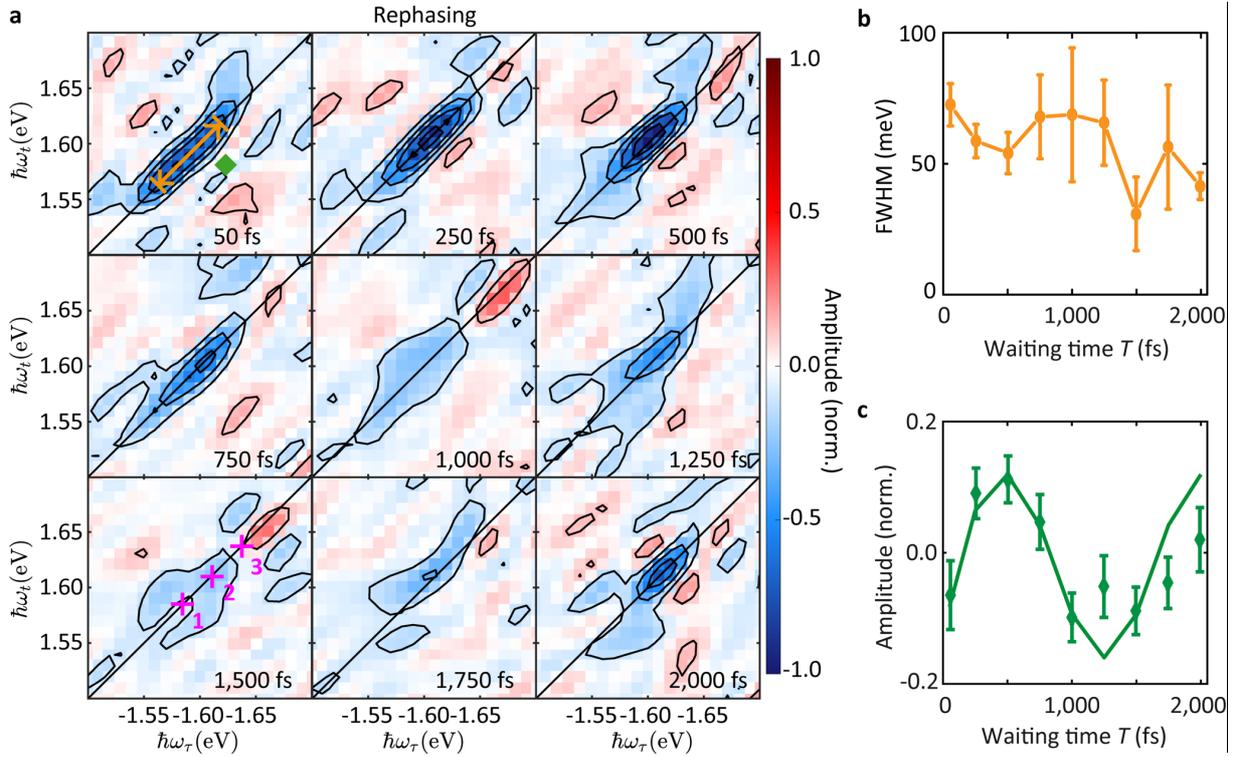

**Figure 2.** Beating signal in the rephasing 2D maps. **a,** Rephasing 2D maps at different $T$, normalized to the maximum absolute value of the real part of the map at $T$ = 500 fs. **b,** Diagonal linewidth (FWHM, indicated by the orange double arrow at $T$ = 50 fs in panel **a**) versus $T$. The error bars depict 95% confidence bounds from fitting the diagonal slices by a Gaussian function. **c,** Amplitude evolution (green diamonds) of one pixel (marked by the green diamond at $T$ = 50 fs in panel **a**) and fit (solid green curve). The error bars are evaluated by calculating the fluctuations within a region containing background noise (Supplementary Note 4).

The rephasing 2D maps in the region around the A exciton are shown in Fig. 2a for various $T$, while the nonrephasing and absorptive 2D maps are in Supplementary Figs. 1 and 5, respectively. The peak linewidth along the diagonal direction of the rephasing map (orange double arrow in upper left panel) is plotted versus $T$ in Fig. 2b. Closer analysis of the systematic variation of this linewidth with $T$ (Supplementary Note 2) indicates that there are 3 components along the diagonal, marked with purple crosses in the lower left panel of Fig. 2a, whose amplitudes oscillate, but not in phase. Thus, when $T$~1,500 fs, the amplitude of the middle component is much higher than the other two, minimizing the effective diagonal linewidth (minimum in Fig. 2b). The measured 2D maps capture the fourth-order nonlinear optical response, as sixth-order contributions are negligible (Supplementary Note 3).

We then extract the amplitude evolution of an exemplary pixel (marked by the green diamond in the 2D map at 50 fs) as a function of $T$ (Fig. 2c). The number of points is restricted due to the long measurement time (26 h for one point). A long-lived (>2 ps) oscillation with amplitude above the noise level is observed. The reproducibility of the data is confirmed by a second measurement for the same $T$ in Supplementary Note 4.

We now analyze the origin of the oscillations in the 2D maps with the goal to deduce the EXPC strength. Previous experiments reported that the trion signal in 1L-MoSe$_2$, located ~0.03 eV below the neutral exciton peak[56], gradually dies out both in PL and absorption when T increases from 15 to 295 K, while the signal intensity of neutral excitons remains nearly unchanged[25,56]. Thus, the signal of the trion is much weaker than that of the neutral exciton at room temperature and in our experiment we detect predominantly the neutral exciton. This implies that wave packets involving trions can be excluded as a source of the long-lived (>2 ps) RT oscillations in Fig. 2c. Biexciton signals can be excluded in our 2D measurements due to their thermal dissociation at RT and cancellation of excited-state absorption pathways in fluorescence-detected 2D spectroscopy (see Supplementary Note 5). Vibrational wave packets were reported at RT in Ref. 21,22, with a dephasing time ~4.5 ps for 1L- and few-layer WSe$_2$[21] and ~1.7 ps for 1L-MoS$_2$[22]. Therefore, EXPC can explain the oscillations in our 2D maps. We extract the phonon energy from a fit (Fig. 2c, solid green curve) and obtain, even for our undersampled (less than one data point for each oscillation period as a result of a compromise arising from finite available data acquisition time) data, an oscillation period ~136±2 fs (see Supplementary Note 6 for the fitting procedure). This corresponds to an energy splitting between the participating states ~30.4±0.4 meV, matching the optical $A'_1$ phonon mode's energy ~29.9 meV, i.e., 241 cm$^{-1}$, as measured in the Raman spectrum of Fig. 1d.

We define the EXPC strength using the Huang–Rhys factor, $S$, in the framework of the Franck–Condon coupling model[57] (see Supplementary Note 7 for a definition of $S$), with the minimum number of states needed to describe the observed data (Fig. 3a). The model of Fig. 3a delivers 3 transition energies, as observed experimentally (purple crosses in Fig. 2a): We assign component 1 (with the lowest energy $\hbar\omega_1$) to the transition between $|g_1\rangle$ and $|e_0\rangle$ (blue color in Fig. 3), component 2 (with a higher energy $\hbar\omega_2$) to the two degenerate transitions between $|g_0\rangle$ and $|e_0\rangle$ and between $|g_1\rangle$ and $|e_1\rangle$ (black and green colors, respectively), and component 3 (with the highest energy $\hbar\omega_3$) to the transition between $|g_0\rangle$ and $|e_1\rangle$ (red color).

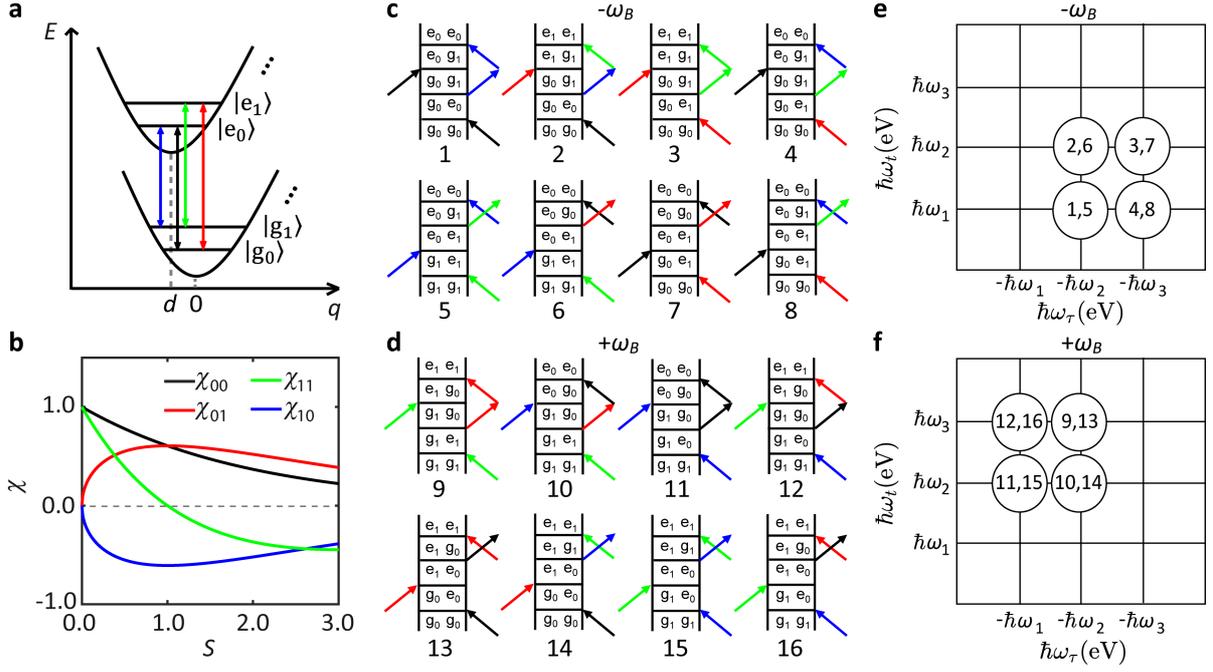

**Figure 3.** Analysis of beating signals. **a,** Schematic diagram of displaced harmonic oscillators (Franck–Condon coupling model[57]) with two vibrational levels ($|g_0\rangle$ and $|g_1\rangle$) in the electronic ground state and two in the electronic excited state ($|e_0\rangle$ and $|e_1\rangle$). The horizontal shift between the two potential minima, $d$, characterizes the exciton–phonon coupling strength. Transitions $|g_0\rangle$–$|e_0\rangle$, $|g_0\rangle$–$|e_1\rangle$, $|g_1\rangle$–$|e_0\rangle$, and $|g_1\rangle$–$|e_1\rangle$ are color-coded in black, red, blue, and green, respectively. **b,** Dependencies of Franck–Condon amplitudes $\chi_{ij}$ ($i, j$ = 0 or 1) on $S$, which scales as $S = d^2/2$. **c,d,** Feynman pathways giving rise to the beating signals with (**c**) negative beating frequency $-\omega_T$ and (**d**) positive frequency $+\omega_T$. **e,** Beating-map locations of numbered Feynman pathways from panel **c**. **f,** Beating-map locations of numbered Feynman pathways from panel **d**.

Transitions between $|g_0\rangle$ and $|e_{i\geq 2}\rangle$ states are not observed in the 2D maps. This agrees with resonance Raman scattering[19,20] and their time-domain analogues[21,22], where the $A'_1$ overtone was not detected. This may imply an efficient nonradiative decay channel for the $|e_2\rangle$ state, which results in a fast dephasing time for the hot vibronic band transitions. Transitions between $|g_{i\geq 2}\rangle$ and $|e_0\rangle$ are also not observed in the 2D maps, which can be explained as a negligible thermal population of $|g_{i\geq 2}\rangle$ due to a small Boltzmann factor at RT. The transition amplitudes between different vibronic sublevels (blue, black, green, and red arrows in Fig. 3a) are proportional to the overlap of the vibrational wave functions of initial and final state, i.e., the Franck–Condon amplitudes $\chi$ [58], plotted as a function of $S$ in Fig. 3b. At $S = 0$, the red and blue curves are zero, indicating that it is not possible to excite $|e_1\rangle$ starting from $|g_0\rangle$ or to reach $|g_1\rangle$ from $|e_0\rangle$, thus the electronic/excitonic excitation is decoupled from vibrations.

We now correlate $S$ with the oscillatory signals. We perform an additional Fourier transformation of 2D maps with respect to $T$. This gives rise to a three-dimensional (3D) spectrum, which is a hypercube as a function of $\hbar\omega_\tau$, $\hbar\omega_T$ and $\hbar\omega_t$. 2D cuts at $\hbar\omega_T = +\hbar\omega_B$ and $\hbar\omega_T = -\hbar\omega_B$ result in two 2D beating maps, where $\omega_B$ is the beating frequency induced by EXPC.

Figure 3c lists all possible rephasing Feynman pathways that can result in contributions at negative beating frequency -$\omega_B$. Their individual positions in the 2D beating maps are in Fig. 3e. Figure 3d contains the contributions at positive $\omega_B$, and Fig. 3f their positions in the 2D map. The determination of all peak positions of individual Feynman pathways in 2D beating maps is introduced in Supplementary Note 8. Adding all pathways, we expect the beating map to be located on the lower right of the diagonal for negative beating frequency (Fig. 3e), and on the upper left for positive (Fig. 3f). The precise shape of the overall beating map depends on the relative amplitudes of the individual Feynman pathways. Those depend on the initial populations of $|g_0\rangle$ and $|g_1\rangle$, hence on the sample temperature, and on the products of the Franck–Condon amplitudes of the 4 involved transitions (colored arrows in Figs. 3c,d) that in turn depend on $S$ (Fig. 3b). Thus, analyzing the shape of the beating maps allows us to estimate $S$.

For a quantitative evaluation, we simulate the 2D beating maps by numerically solving a Lindblad master equation[59] for a system described by the Franck–Condon model illustrated in Fig. 3a (see Methods for details). $S$ is varied from 0.25 to 2 with a step size of 0.25. Fig. 4a plots the simulation for $S$ = 0.5, 1, and 1.5 from top to bottom. Data for other $S$ are in Supplementary Fig. 13. We recognize the expected features of Figs. 3e,f. The pathway contributions overlap with each other, due to line broadening along the diagonal and anti-diagonal directions. For $S$ = 1.5, the 4 underlying subpeaks create a square lineshape. For smaller $S$, the anti-diagonal linewidth changes strongly because of the varying relative contributions of the different Feynman pathways, leading to one asymmetric peak in each 2D beating map, whose center is located below (above) the diagonal line for negative (positive) beating frequency as predicted in Fig. 3e (Fig. 3f). The change in linewidth can be understood by considering that $\chi_{11}$ (Fig. 3b, solid green curve) crosses zero (the dashed gray line) for $S$ = 1, such that only Feynman pathways 1, 7, 11, 13, e.g., without $|g_1\rangle$–$|e_1\rangle$ transition (green arrow in Figs. 3c,d), contribute. Therefore, the anti-diagonal linewidth reaches a minimum for $S$ = 1.

Fig. 4b shows the experimental 2D beating maps at $-\omega_B$ (left) and $+\omega_B$ (right), obtained as cuts through the rephasing 3D spectrum at the same beating frequency as in the simulations, $\omega_B = 4.6 \times 10^{13}$ s$^{-1}$. The asymmetry with respect to the diagonal is visible, and the elliptical shape [rather than roundish (small $S$) or squarish (large $S$)] points at an intermediate $S$ by comparison with simulations. The lowest contour lines of the experimental and the simulated beating maps in Figs. 4a,b show some "jagged" behaviour. The factors that could contribute to this are discussed in Supplementary Note 9.

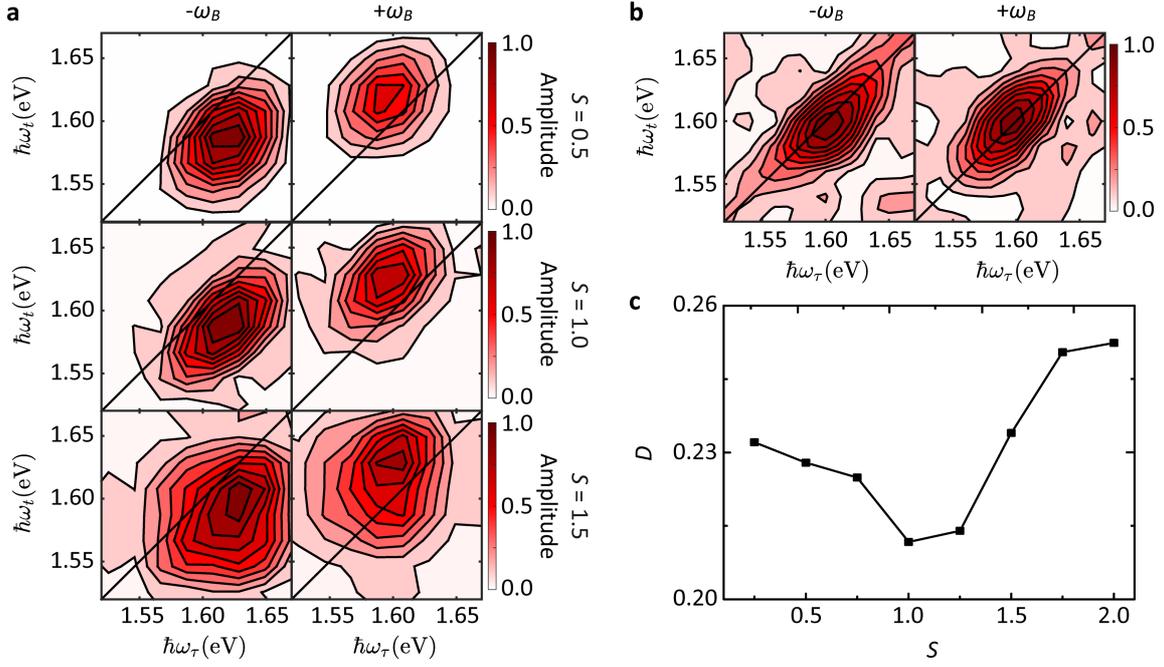

**Figure 4.** 2D beating maps. **a,** Simulated 2D beating maps for $-\omega_B$ (left) and $+\omega_B$ (right) and $S$ = 0.5, 1, 1.5 from top to bottom rows). **b,** Measured 2D beating maps with $-\omega_B$ (left) and $+\omega_B$ (right). **c,** The deviation, $D$, between measured and simulated 2D beating maps versus $S$ used in the simulation.

To determine the EXPC strength quantitatively, we calculate the deviation $D$ between measured and simulated 2D beating maps:

$$D = \sqrt{\frac{1}{N^2}\sum_{i=1}^{N}\sum_{j=1}^{N}\left(A_{ij}-\tilde{A}_{ij}\right)^2}, \qquad (1)$$

where $N$ is the pixel number in each dimension of the 2D beating maps, $A_{ij}$ ($\tilde{A}_{ij}$) is the amplitude of the pixel in column $i$ and row $j$ of the simulated (experimental) 2D beating map. Figure 4c plots $D$ versus $S$. We find the best agreement for $S = 1$. We then compare the experimental regular absorptive, rephasing, and nonrephasing 2D maps for $T = 50$ fs (Fig. 5a) with the simulation using the optimal $S$ (Fig. 5b) and find good agreement, confirming the reliability of our Franck–Condon model.

We note that large Huang–Rhys values, $S$, on the order of 1 in 1L-TMDs are supported by theory[60–62], but were never previously experimentally measured, to the best of our knowledge.

The exciton coupling with longitudinal optical phonons in 1L-TMDs was studied by *ab initio* calculations[60,61]. These found that polar LO phonon vibrations give rise to a macroscopic electric field that couples to the charge carriers. Such a coupling, named "Fröhlich interaction", is fundamentally affected by the dimensionality of the system. When the dimensionality of the system decreases from 3D to 2D, a 3-fold increase of Huang–Rhys factor is predicted, see, e.g., Fig. 7 in Ref. 63. Taking into account Fröhlich interaction in a 2D model, Ref. 62 calculated $S$ for LO phonons as a function of the polarization parameter for 1L-MoSe$_2$, finding ~1.93–2.24. Defects may also have a strong influence on the Huang–Rhys factor $S$[64,65]. The electric fields induced by local charges at interfaces increase the non-vanishing part of the electron and hole polaron clouds in the exciton state[65] and as a result, $S$ as large as ~1 can be found[65].

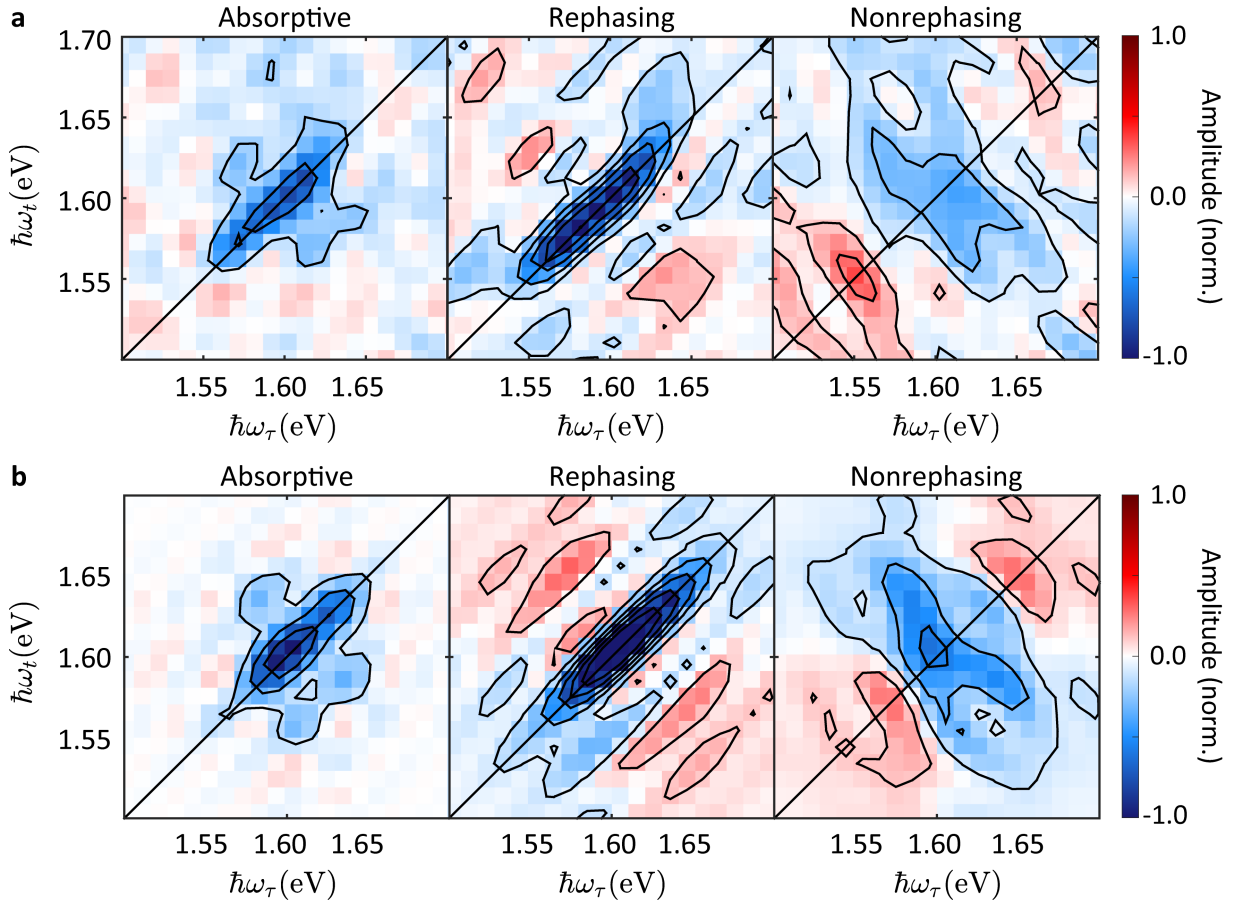

**Figure 5.** Absorptive (left), rephasing (middle), and nonrephasing (right) real-valued 2D maps at $T$ = 50 fs. **a,** Experiment. **b,** Simulation using the deduced optimal $S$ = 1.

We carried out spatially resolved, fluorescence-detected 2D micro-spectroscopy on 1L-MoSe$_2$. We identified phonon sidebands upon excitation of the A exciton, due to coupling to the optical phonon mode $A'_1$. While the phonon is not resolved in linear absorption or PL spectra at room temperature, analysis of the 2D beating frequency as a function of waiting time allowed us to assign the phonon mode via comparison with Raman data. We determined the exciton–phonon coupling strength, i.e., the displacement along the phonon coordinate of the excited-exciton oscillator potential with respect to the ground state, and found a Huang–Rhys factor, $S\sim1$, by comparison with simulations of 2D beating maps. The measured $S\sim1$ is larger than most reported values ($S\sim0$–$0.5$) of other inorganic semiconductor nanostructures[53], such as CdSe quantum dots[49] and rods[50], ZnSe quantum dots[51], single-wall carbon nanotubes[52], etc., indicating a strong EXPC. This finding may benefit, among others, the development of TMD-based polariton devices[66], in which the polariton-relaxation process strongly depends on the EXPC strength[67].

Our space-, time-, and excitation/detection-frequency-resolved spectroscopy provides a unique tool to measure EXPC strength also in other layered TMDs. hBN encapsulation can lower inhomogeneous broadening of 1L-TMDs[68,69], we thus expect better resolved peaks for hBN-encapsulated samples. However, the Huang–Rhys factor, $S$, may be influenced by the substrate by changing the macroscopic electric field induced by the polar LO phonon at the interface[61]. E.g., SiO$_2$ increases the screening of the Fröhlich interaction strongly at small momenta[61]. Therefore, we expect that a different substrate might result in a different Huang–Rhys factor, hence, hBN encapsulation might also influence $S$. Our method can be extended to other semiconducting systems for which phonon-induced subbands are expected in the excitonic lineshape, such as single-wall carbon nanotubes[70], layered perovskites[71], bulk heterojunctions[72], or other organic crystals. Because of the high spatial resolution of ~260 nm, our technique can also be used to study excitonic coupling in layered materials heterostructures or microcavities with embedded semiconductors. The determination of EXPC will provide design-relevant parameters for the development of photonic and optoelectronic devices based on these semiconducting systems.

**Methods**

**Samples fabrication.** The samples are prepared by micromechanical cleavage[73] of bulk $MoSe_2$ from HQ Graphene. Micromechanical cleavage is performed with polydimethylsiloxane (PDMS) and, after inspection under an optical microscope, 1L-$MoSe_2$ is dry transferred in ambient conditions to a 200 μm fused silica substrate[74]. After transfer, the samples are characterized by Raman and PL with a Renishaw Invia spectrometer at 514 nm and with a 50× objective. Metallic frames (Cr/Au) are fabricated around selected 1L-$MoSe_2$ flakes on fused silica by laser-writer lithography to facilitate the identification of the samples' position.

**Data acquisition.** A femtosecond oscillator (Venteon Laser Technologies GmbH, Pulse One PE) provides a laser spectrum ranging from 650 to 950 nm, confined by a hard aperture in the Fourier plane of a 4*f*-based pulse shaper in front of the liquid-crystal display (LCD, Jenoptik Optical Systems GmbH, SLM-S640d). The aperture acts as a short-pass filter at 808 nm, so that the longer-wavelength PL can be detected without scattering from the pump light. A Schott KG5 color filter further modulates the spectrum into a smooth shape, which ensures the absence of pronounced side peaks and other irregularities in the temporal pulse profile. The laser focus in the microscope is mapped by a piezo scanning stage (P-517.3CL, PI, Germany). Excitation occurs through a focusing objective (Nikon Plan Apo, 100×/1.40). PL is collected through the same objective, transmitted through a dichroic beam splitter (DBS, AHF Analysentechnik, F48-810) and an additional emission filter (EF, AHF Analysentechnik, F76-810), and detected by an APD (Perkin Elmer, SPCM-CD 2801).

We compress the laser pulses by (1) using chirped mirrors to pre-compensate some amount of second-order phase dispersion; (2) employing the pulse shaper to compensate any remaining dispersion. A two-photon photodiode (TPPD) is placed in the focus of the microscope objective to generate a nonlinear feedback signal that is a measure of pulse intensity and pulse duration. We then utilize the algorithm presented in Ref. 75 to maximize the peak intensity, leading to a transform-limited laser pulse. To characterize the result of pulse compression, an autocorrelation trace is measured using the same TPPD, Supplementary Figure 15. This agrees well with a simulated one assuming the experimentally measured laser spectrum and a flat spectral phase. This correspondence indicates successful phase-dispersion compensation and ~12 fs pulses at the sample position, as discussed in Refs. 48,75.

Linearly polarized light, acting as a superposition of left- and right-handed circularly polarized light, is used to simultaneously excite both the transitions in the K and K' valleys. The pump fluence is ~2 µJ/cm². We estimate the heating through laser irradiation during the experiment as discussed in Supplementary Note 10. The sample temperature increases from 300 to ~308 K within the first 100 ns, then remains constant, thus there is no unwanted heating of the sample, thermal instabilities or damage.

We obtain the 2D maps by scanning $\tau$ and $t$ in steps of 3 fs each from 0 to 99 fs, for $T$ = 50, 250, 500, 750, 1000, 1250, 1500, 1750, 2000 fs, using the spectral modulation function[76]:

$$M(\omega) = \exp[i(\omega - \omega_0(1-\gamma))(-\tau-T)] + \exp[i(\omega - \omega_0(1-\gamma))(-T) + i\varphi_{12}] + \exp[i\varphi_{13}] + \exp[i(\omega - \omega_0(1-\gamma))t + i\varphi_{14}], \qquad (2)$$

at a center frequency $\omega_0 = 2.5 \times 10^{15}$s$^{-1}$. We avoid undersampling with time steps of 3 fs by employing a partially rotating frame with $\gamma$ = 0.2. The third pulse is fixed at time 0, so that when 2D maps are measured at a certain $T$, only the first and fourth pulses are delayed. By setting the phase of the first pulse to 0, three relative phases, i.e., $\varphi_{12}$, $\varphi_{13}$, and $\varphi_{14}$, are scanned in a 27-step phase-cycling scheme, where each relative phase takes values of $0$, $\frac{2\pi}{3}$, and $\frac{4\pi}{3}$. This allows us to select rephasing and nonrephasing contributions individually from the complete raw data[77,48]. We obtain absorptive 2D maps by summing the real parts of the rephasing and nonrephasing 2D maps, cancelling dispersion terms, leaving a pure absorptive lineshape[40]. Due to the finite response time of the liquid crystals of our pulse shaper, we wait ~500 ms after changing the phase mask before taking data. PL is averaged over ~1 ms for each APD acquisition period. Including additional averaging (2000 times for each pulse shape), the total measurement time for one 2D map is ~26 h. During the measurements the PL intensity of the sample is constantly monitored every ~80 s. We observe no systematic decay during the measurement time. This indicates a long-term chemical, thermal, and photostability of the sample. The group delay dispersion at the sample position is compensated by adding an additional phase to the modulation function[48].

**Simulations.** To simulate the 2D maps, we solve the Lindblad quantum master equation[59]

$$\frac{\partial}{\partial t'}\rho(t') = -\frac{i}{\hbar}[\mathcal{H}(t'),\rho(t')] + \sum_j \frac{1}{T_j}\left(\mathcal{L}_j\rho(t')\mathcal{L}_j^\dagger - \frac{1}{2}\mathcal{L}_j^\dagger\mathcal{L}_j\rho(t') - \frac{1}{2}\rho(t')\mathcal{L}_j\mathcal{L}_j^\dagger\right), \qquad (3)$$

where the time evolution of the density matrix $\rho(t')$ of the quantum system under a Hamiltonian $\mathcal{H}(t')$ is treated in the Liouville–von Neumann formalism, with the extension of dissipative and pure dephasing effects, $\mathcal{H}(t')$ is expressed as the sum of a time-independent Hamiltonian $\mathcal{H}_0 = \hbar\omega_m\sum_m|m\rangle\langle m|$ and an interaction Hamiltonian $\mathcal{H}_I(t') = \gamma_{\text{ex}} E(t')\sum_{m\neq n}\mu_{m,n}(|m\rangle\langle n| + |n\rangle\langle m|)$, where $|m\rangle$ (or $|n\rangle$) are the unperturbed eigenstates with eigenenergies $\hbar\omega_m$ (or $\hbar\omega_n$), $\gamma_{\text{ex}}$ is the field coupling strength for excitation with the external electric field $E(t')$, $\mu_{m,n}$ is the transition dipole moment between states $|m\rangle$ and $|n\rangle$, $T_j$ represents the time associated with a pure dephasing or population relaxation process, and the Lindblad operators $\mathcal{L}_j$ are defined as $\mathcal{L}_j = a_n^\dagger a_n$ for pure dephasing and $\mathcal{L}_j = a_m^\dagger a_n$ with $m \neq$ $n$ for a population relaxation process, where $a_m^\dagger$ and $a_n$ denote the creation and annihilation operators, respectively.

We assume a four-level system, with two vibrational levels in the ground electronic state ($|g_0\rangle$ and $|g_1\rangle$) and two vibronically excited states ($|e_0\rangle$ and $|e_1\rangle$), as in Fig. 3a. The splittings within the subbands are taken to be identical, i.e., we use the same energy separations (30 meV[21]) between $|g_0\rangle$ and $|g_1\rangle$ as well as between $|e_0\rangle$ and $|e_1\rangle$. The Franck–Condon amplitudes between $|g_i\rangle$ and $|e_j\rangle$, i.e., $\chi_{ij}$ ($i,j$ = 0 or 1) depend on $S$ as for Fig. 3b. The initial populations of $|g_0\rangle$ and $g_1\rangle$ are determined by the temperature, according to the Boltzmann distribution. In Supplementary Note 7 we estimate the heating through laser irradiation during the experiment and find the sample to remain close to RT.

The excitation laser field is calculated from the experimentally utilized laser spectrum assuming a flat phase and then adding the transfer function:

$$M(\omega) = \exp[i(\omega - \omega_0(1-\gamma))(T_{\text{off}})] + \exp[i(\omega - \omega_0(1-\gamma))(T_{\text{off}} + \tau) + i\varphi_{12}] + \exp[i(\omega - \omega_0(1-\gamma))(T_{\text{off}} + \tau + T) + i\varphi_{13}] + \exp[i(\omega - \omega_0(1-\gamma))(T_{\text{off}} + \tau + T + t) + i\varphi_{14}], \quad (4)$$

where $T_{\text{off}}$ is an offset of the position of the first pulse in time domain, set at 100 fs to avoid cutting off the first pulse at time zero. In the experimental modulation function of Eq. 2, time zero is set at the maximum of the third pulse, leading to a different mathematical expression. However, this difference does not affect the resulting 2D maps, since only relative time delays between the pulses are relevant. $\tau$ and $t$ in the simulation are scanned with the same parameters as in the experiment, from 0 to 99 fs in steps of 3 fs with $\gamma = 0.2$, whereas $T$ is scanned from 0 to 200 fs in steps of 25 fs.

Inhomogeneous broadening due to a Gaussian distribution of excitonic transition frequencies is taken into account by obtaining the inhomogeneously broadened response function, $S_I(\tau, t)$, from the homogeneous response, $S(\tau, t)$ from solving Eq. 3, via:

$$S_I(\tau, t) = S(\tau, t) \cdot \exp[-\Delta^2 \cdot (\tau \mp t)^2], \tag{5}$$

where $\Delta$ is a parameter linearly proportional to the inhomogeneous linewidth broadening, - is applied for the rephasing signal, and + for the nonrephasing signal. Eq. 5 is used under two assumptions. 1) Spectral diffusion can be ignored within the $T = 2$ ps window of the measurements. Typically, spectral diffusion is caused by environmental fluctuations around the transition dipoles, inducing a broadening along the anti-diagonal direction for the absorptive 2D maps as $T$ increases[40]. This is not observed in our experiments (Supplementary Figure 5), indicating a much slower than 2 ps modulation time constant of the environment, justifying the use of Eq. 5. 2) The vibrational frequency does not change with the excitonic transition energy, also assumed for the model of Fig. 3a and Eqs. 1–3. If this was not fulfilled, a tilt of elongated peaks in the 2D beating maps relative to the diagonal would be observed[41], unlike in our measurements (Fig. 4b).

## Data availability

The data that support the findings of this study have been deposited in Mendeley Data with the embargo date of March 1, 2021. The data are available at the following link:

http://dx.doi.org/10.17632/52yj8d9p4t.1

## Acknowledgements

We thank Pavel Malý for support with computations of 2D maps. We acknowledge funding by the European Research Council (ERC) – Grants No. 614623, Hetero2D, GSYNCOR; EPSRC

Grants EP/K01711X/1, EP/K017144/1, EP/N010345/1, and EP/L016087/1; and the EU Graphene and Quantum Flagships.

## Author contributions

D.L., C.T., and M.N. performed spectroscopic measurements. D.L. analyzed the 2DES data and conducted the 2DES simulations. S.D.C. supervised C.T.; T.B., G.C. and A.C.F. initiated and supervised the project. G.S. and G.W. prepared and characterized the samples. All authors discussed results. D.L., C.T., G.C., T.B. and A.C.F. wrote the paper with input from all coauthors.

## Competing interests

The authors declare no competing interests.

## Additional information

**Supplementary information** is available for this paper.

# Supplementary Information:

# Exciton–phonon coupling strength in single-layer MoSe$_2$ at room temperature

*Donghai Li[1], Chiara Trovatello[2], Stefano Dal Conte[2], Matthias Nuß[1], Giancarlo Soavi[3,4], Gang Wang[3], Andrea C. Ferrari[3*], Giulio Cerullo[2,5*], and Tobias Brixner[1,6*]*

[1]Institut für Physikalische und Theoretische Chemie, Universität Würzburg, Am Hubland, 97074 Würzburg, Germany

[2]Dipartimento di Fisica, Politecnico di Milano, Piazza L. da Vinci 32, I-20133 Milano, Italy

[3]Cambridge Graphene Centre, University of Cambridge, 9 JJ Thomson Avenue, Cambridge CB3 0FA, UK

[4]Institute for Solid State Physics, Abbe Center of Photonics, Friedrich-Schiller-University Jena, Max-Wien-Platz 1, 07743, Jena, Germany

[5]IFN-CNR, Piazza L. da Vinci 32, I-20133 Milano, Italy

[6]Center for Nanosystems Chemistry (CNC), Universität Würzburg, Theodor-Boveri-Weg, 97074 Würzburg, Germany

*e-mails: acf26@eng.cam.ac.uk, giulio.cerullo@polimi.it, brixner@uni-wuerzburg.de

# Supplementary Note 1: Rephasing and nonrephasing 2D maps

In collinear 2DES, rephasing and nonrephasing signals can be obtained separately by phase cycling[1–3]. The rephasing 2D maps of 1L-MoSe$_2$ at different $T$ are in Fig. 2a in the main text, Supplementary Fig. 1 plots the nonrephasing 2D maps at different $T$. All the rephasing and nonrephasing 2D maps are normalized to the maximum absolute value of the real part of the rephasing maps at $T$ = 500 fs.

In order to better understand the origin of the coherent oscillations in the measured 2D maps, we summarize some properties of double-sided Feynman diagrams[4]. We show exemplary Feynman diagrams for a three-level system (i.e., without vibrational sublevels within electronic states) of rephasing and nonrephasing pathways for population-detected 2DES (Supplementary Fig. 2, obtained via measuring fluorescence in collinear geometry) as well as for conventional coherence-detected 2DES (Supplementary Fig. 3, obtained in partially non-collinear geometry). The vertical lines represent time evolution of ket (left) and bra (right) states of the density matrix with time running from the bottom to the top. Every pulse induces an impulsive transition of the system within the density-matrix description[4], depicted by a solid arrow. An arrow pointing towards the quantum states in the middle increases the respective electronic quantum number, an arrow pointing away decreases it (|g⟩ denotes the ground state, |e⟩ the first excited state, and |f⟩ the second excited state). Between two impulsive transitions, the system either remains in a population state (identical bra and ket states) or propagates in a coherent state (different bra and ket states)[4]. The sign of the signal of a specific Feynman pathway (labeled on the top of the Feynman diagram) depends on the number of interactions from the right side of the Feynman diagram, wherein each interaction leads to multiplication by -1 of the signal[4].

In all cases, the system is in a coherent state between the first and the second pulse and between the third and the fourth pulse, i.e., the bra and ket states are different from each other. We call the associated time intervals excitation time (or first coherence time) $\tau$ and detection time (or second coherence time) $t$. Two-dimensional Fourier transformation with respect to these two coherence times results in a 2D map with the two axes of excitation frequency $\omega_\tau$ (or energy $\hbar\omega_\tau$) and detection frequency $\omega_t$ (or energy $\hbar\omega_t$)[4].

The sign of the frequency of a coherent state |X⟩⟨Y| is defined, without loss of generality, to be positive if level |X⟩ is higher in energy than level |Y⟩, and negative if level |X⟩ is lower. The exemplary Feynman diagrams reveal that, in the rephasing pathways, the coherent state evolution during $\tau$ is associated with a negative frequency, because the first interaction with a light field occurs from the right side, bringing the bra state from ⟨g| to ⟨e|, while the coherent state during $t$ evolves with positive frequency, because, in that case, the ket state has higher energy than the bra state. The different sign for the temporal evolution in the two coherence times is the reason why this is called the "rephasing" signal, since the phase evolution during the first period is reversed in the second. Therefore, in rephasing maps, $\hbar\omega_\tau$ assumes negative values[4]. By contrast, in nonrephasing pathways, the coherent states during both $\tau$ and $t$ have positive frequencies.

We now consider the differences between the two types of 2DES from the Feynman diagrams in Supplementary Figs. 2,3. For the population-detected approach (Supplementary Fig. 2), there are four impulsive transitions, generating an excited population state (i.e., identical bra and ket states), followed by spontaneous emission of fluorescence. For the coherence-detected approach (Supplementary Fig. 3), three laser pulses are used to create a third-order polarization (i.e., different bra and ket states) which emits a coherent signal in the phase-matched direction. Therefore, when measuring rephasing and nonrephasing 2D maps, population-detected 2DES probes fourth-order nonlinear signals, whereas in the coherence-detected geometry one records the third-order response of the system. The apparent discrepancy in the nonlinearity comes about because in the standard formulation of coherence-detected 2DES the light field is treated classically and the final interaction (dashed arrows) is not counted towards the order of nonlinear response[4]. Third-order coherently detected 2D spectra and fourth-order population-detected 2D spectra can both be described with one and the same underlying generalized fourth-order response function, discussed in Ref. 5.

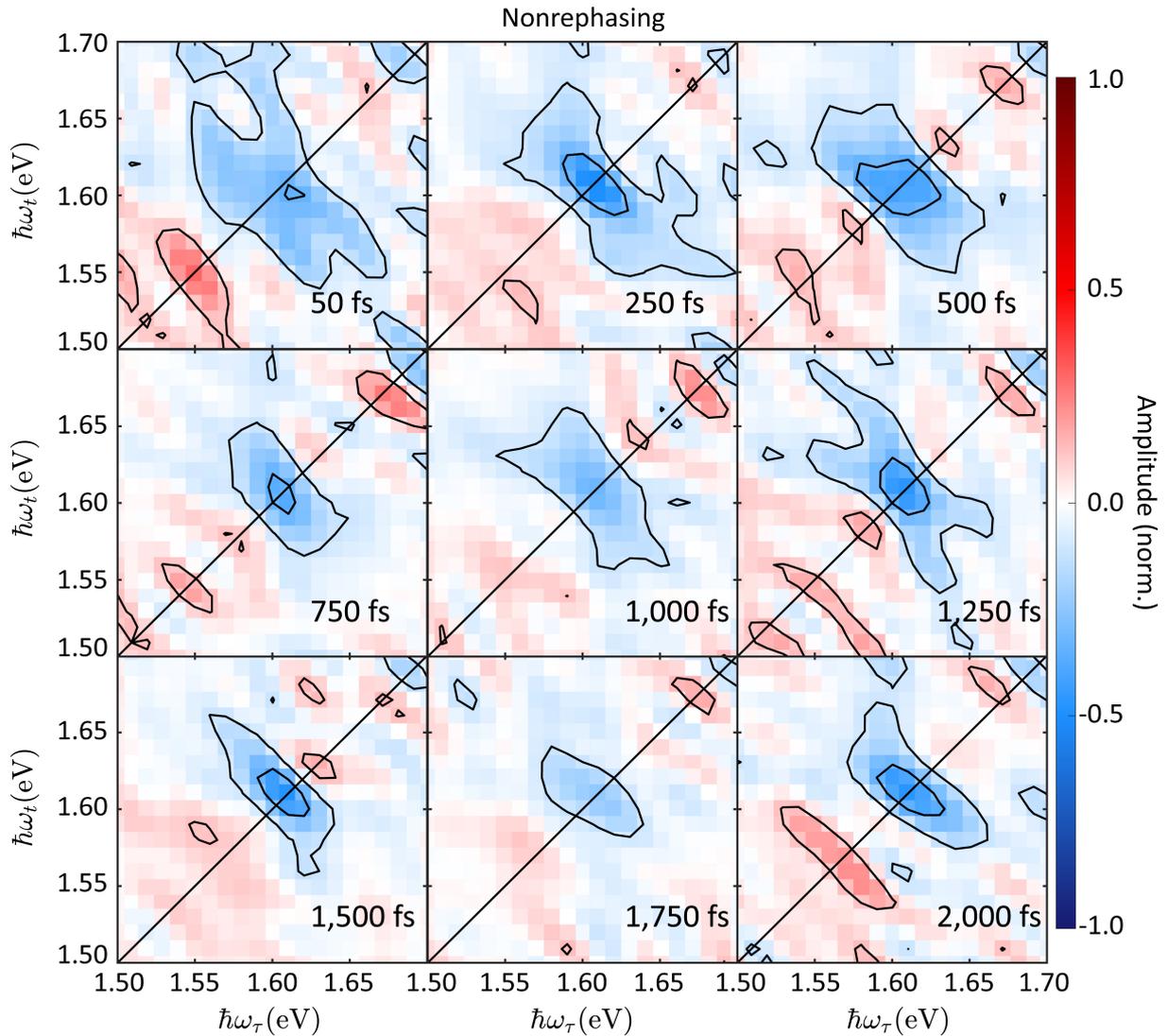

**Supplementary Figure 1:** Nonrephasing 2D maps (real parts) at different *T*.

In coherence-detected 2DES there are three types of Feynman pathways[4] contributing to rephasing and nonrephasing signals, named ground-state bleach (GSB), stimulated emission (SE), and excited-state absorption (ESA), as labeled in Supplementary Fig. 3, wherein the GSB and the SE pathways have positive sign, and the ESA negative. Conversely, in population-detected 2DES, the GSB and SE signals are negative (Supplementary Fig. 2), since there is always one more interaction from the right side compared to the coherence-detected variant. For the ESA signal of population-detected 2DES maps, two pathways exist (ESA 1 and ESA 2) ending up in $|e\rangle\langle e|$ and $|f\rangle\langle f|$ excited population states[6]. If the system arrives at the $|f\rangle\langle f|$ state, fast internal conversion will lead to the $|e\rangle\langle e|$ state before spontaneous emission can occur (Kasha's rule[7]). Hence, ESA 1 and 2 have the same intensities, but opposite signs, under the condition of a unity quantum efficiency for the internal conversion process from $|f\rangle\langle f|$ to $|e\rangle\langle e|$ (i.e., if all the population of the $|f\rangle\langle f|$ state is transferred to the $|e\rangle\langle e|$ state). This results in a cancellation between the two ESA pathways in population-detected 2DES[6]. Such a cancellation is fulfilled here because, if ESA 2 would not fully cancel ESA 1, a left-over ESA signal would appear on the 2D maps at $\hbar\omega_\tau = \hbar\omega_{eg}$ and $\hbar\omega_t = \hbar\omega_{fe}$, where $\hbar\omega_{eg}$ ($\hbar\omega_{fe}$) is the transition energy between $|g\rangle$ and $|e\rangle$ ($|e\rangle$ and $|f\rangle$). Thus, henceforth all the ESA pathways are neglected. For molecular systems, the situation may be different, and one has to include a suitable less-than-unity quantum efficiency parameter and take into account ESA pathways[6,8,9].

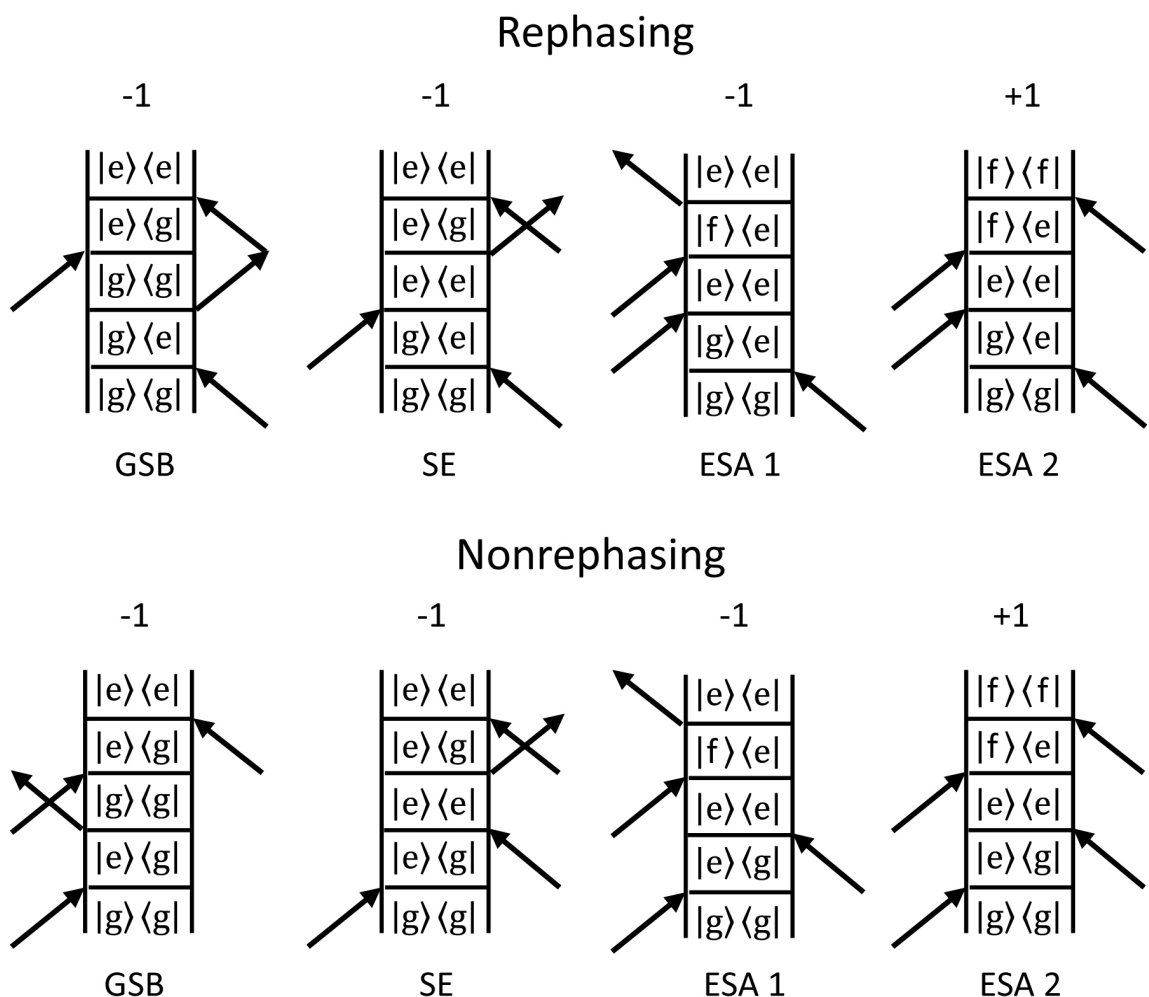

**Supplementary Figure 2:** Typical Feynman diagrams (without considering sublevels within electronic states) of rephasing (top row) and nonrephasing (bottom row) pathways in population-detected 2DES. There are four types of Feynman pathways contributing to rephasing and nonrephasing signals, named ground-state bleach (GSB), stimulated emission (SE), and excited-state absorption (ESA) 1 and 2.

## Supplementary Note 2: Absorptive 2D maps and linewidth analysis

Upon Fourier transformation, both rephasing and nonrephasing signals have dispersive contributions[4]. In order to compare to static or transient absorption spectra, 2D rephasing and nonrephasing maps are normally summed up to cancel the dispersion and obtain an absorptive lineshape[4], as illustrated in Supplementary Fig. 4. The resulting absorptive 2D maps at different $T$ are in Supplementary Fig. 5, normalized to the maximum absolute value at $T$ = 500 fs.

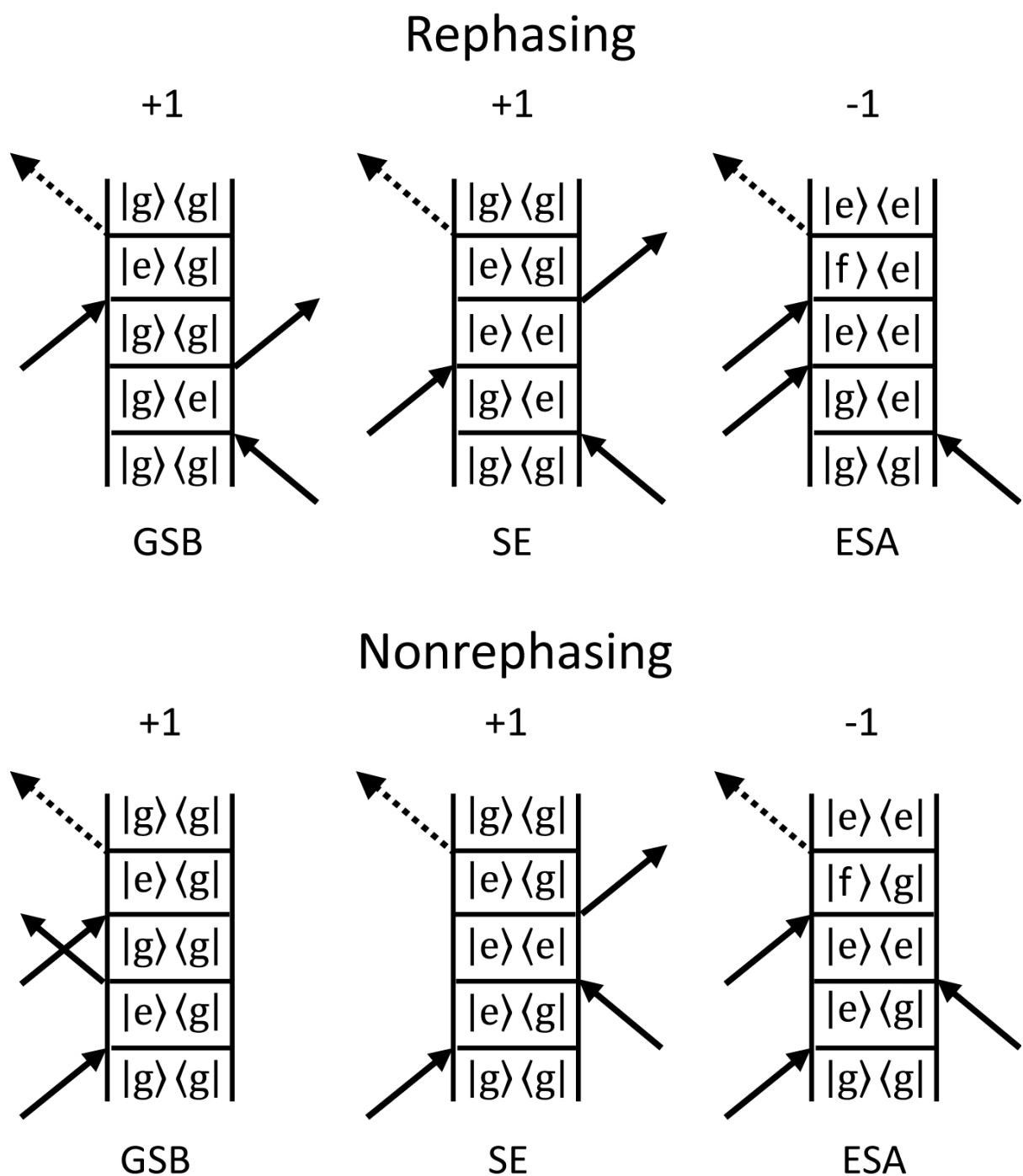

**Supplementary Figure 3:** Typical Feynman diagrams (without considering sublevels within electronic states) of rephasing (top row) and nonrephasing (bottom row) pathways in conventional coherence-detected 2DES. There are three types of Feynman pathways contributing to rephasing and nonrephasing signals, named ground-state bleach (GSB), stimulated emission (SE), and excited-state absorption (ESA).

Absorptive 2D maps simplify the interpretation and allow direct comparison with traditional transient absorption spectroscopy[4]. For studying the origin of coherent oscillations as a function of $T$, it is preferable to analyze separately the amplitude evolutions in the rephasing and nonrephasing maps, because they display different oscillating behaviors for diagonal versus cross peaks. The summation of rephasing and nonrephasing signals will mix these different oscillations and obscure the analysis[10].

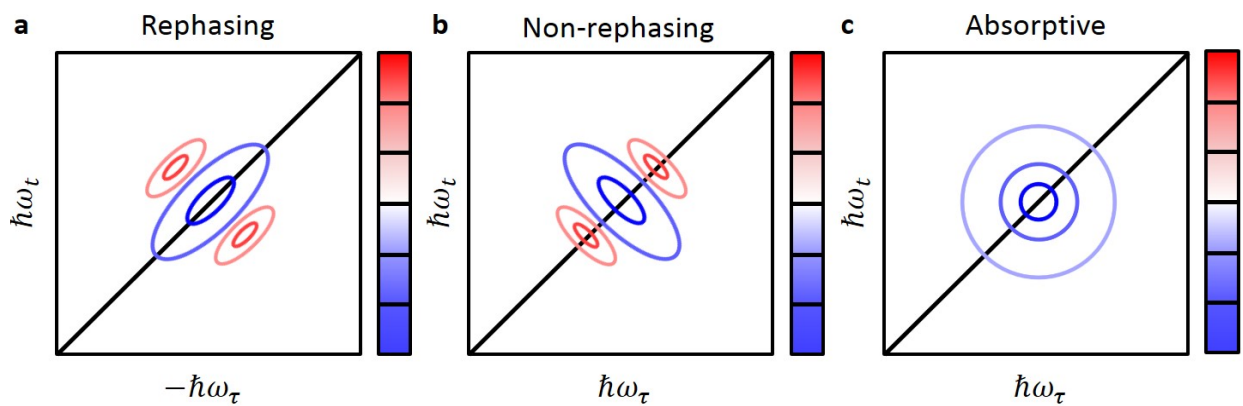

**Supplementary Figure 4:** Schematic real parts of **a**, rephasing, **b**, nonrephasing, and **c**, absorptive 2D maps in the case of a single relevant transition. Both rephasing and nonrephasing signals have dispersive contributions. The summation of their real parts can remove the dispersion, resulting in a purely absorptive map.

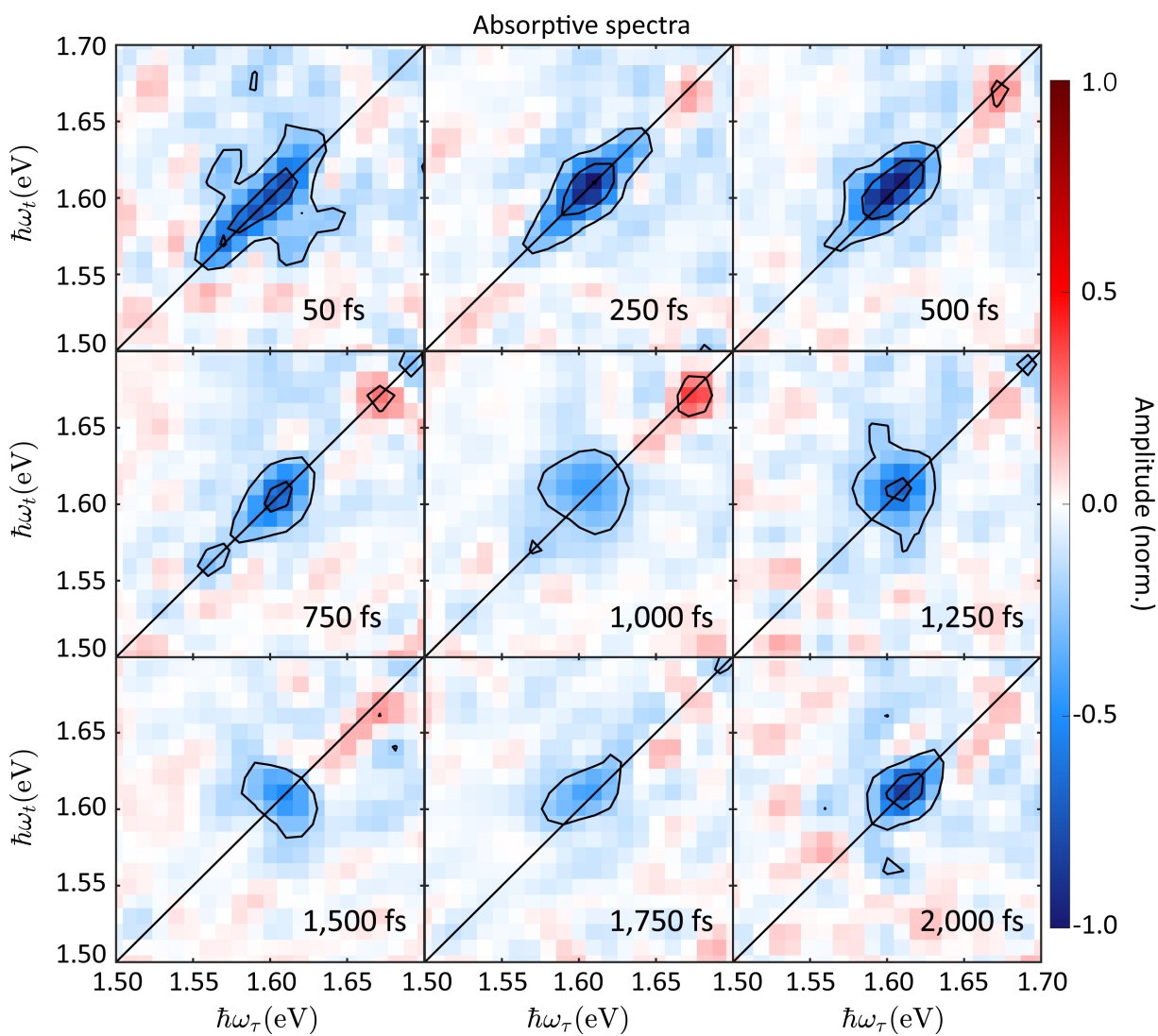

**Supplementary Figure 5:** Absorptive 2D maps at different $T$ normalized to the maximum at $T = 500$ fs.

In the absorptive 2D maps (Supplementary Fig. 5), the linewidth of the peak along the diagonal direction oscillates with $T$ (Supplementary Fig. 6a, black curve). The orange curve in Supplementary

Fig. 6a is the diagonal linewidth of the rephasing maps (the same curve as in Fig. 2b in the main text), which shows the same tendency as the black curve, indicating that there are multiple diagonal components. The amplitudes of these components oscillate, yet not in phase. Supplementary Fig. 6b (6c) shows the absorptive map at $T$ = 250 fs (1500 fs). By comparing the linewidth in diagonal direction (Supplementary Fig. 6d, solid red line for $T$ = 250 fs and solid green line for $T$ = 1500 fs), we find that at 1500 fs the amplitudes of components 1 and 3 drop to zero leaving only component 2, due to the out-of-phase oscillation.

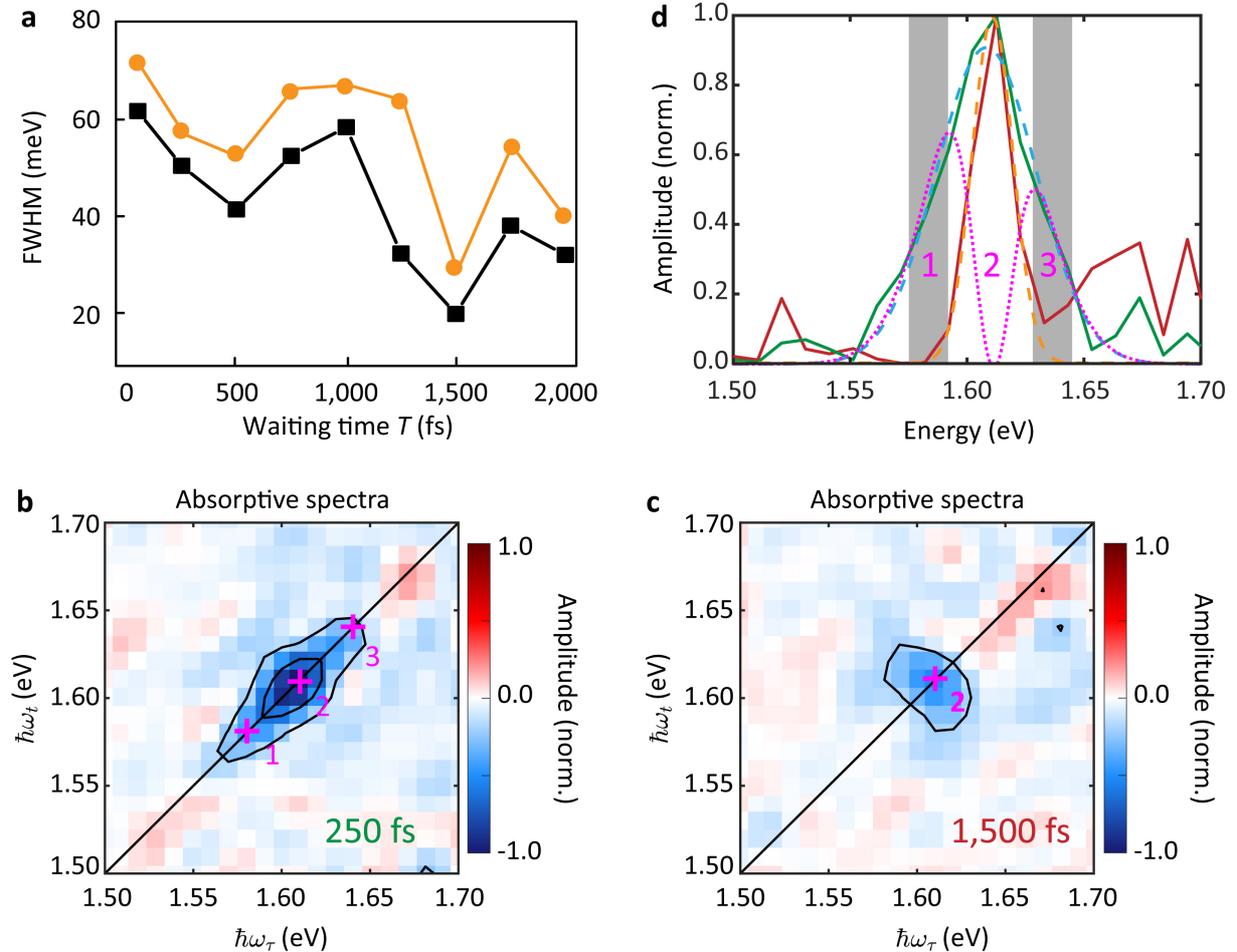

**Supplementary Figure 6:** Line-width analysis. **a,** FWHM along diagonal direction as a function of $T$ from Gaussian fitting of each time step for the absorptive (black curve) and rephasing (orange curve) 2D maps. **b**, Absorptive 2D map at $T$ = 250 fs. **c**, Absorptive 2D map at $T$ = 1500 fs. **d**, Slices along diagonal direction for the 2D maps of panels b (solid green) and c (solid red), their Gaussian fitting curves (dashed orange and dashed green, respectively), as well as the difference of the two Gaussian curves (dotted purple). The gray areas mark the estimated ranges of the center positions of components 1 and 3 as labeled.

Both linewidths are fitted by Gaussians (Supplementary Fig. 6d, dashed orange line for $T$ = 250 fs and dashed cyan-blue line for $T$ = 1500 fs). The FWHM of component 2 is 20±8 meV, smaller than the FWHM ~50 meV derived from linear absorption at RT[11], because the peak intensity in 2D map is proportional to the fourth power of the transition dipole strength[1], while linear absorption scales quadratically with transition dipole strength[4], resulting in a twice-smaller Gaussian linewidth in 2D maps than in linear ones. The laser spectrum additionally modulates the peak shape, and influences the comparison between the linewidth extracted from 2D maps and absorption spectra.

The position of component 2 is ~1.611±0.004 eV, as determined from the Gaussian fit at T = 1,500 fs, while the approximate positions of components 1 and 3 can be deduced from the difference of the two Gaussian fitting functions at time delays of 250 and 1,500 fs (Supplementary Fig. 6d, dotted purple line). The two maximum positions in the difference spectrum are 1.591 and 1.629 eV. However, in the laser spectrum (Supplementary Fig. 7), there is a structured peak with maximum position~1.620 eV. Such a structure will shift the peak positions of different components towards the (local) laser peak maximum position (1.620 eV). Thus, the peak of component 1 will be blue-shifted and component 3 red-shifted. For this reason, we cannot obtain precise center positions of components 1,3 just from maximum peak positions. We estimate their range by setting 1.591 eV (1.629 eV) as the upper (lower) limit of the position of component 1 (component 3), and the position at half the maximum on the low-energy (high-energy) side as the lower (higher) limit. The estimated ranges of the two components are marked by gray areas in Supplementary Fig. 6c. The energy splitting between components 1 and 2 falls in the range ~20–35 meV and between 2 and 3 in the range ~18–33 meV. The estimated center positions of the diagonal peaks of components 1 and 3 are determined by taking the mid-point of each range, which are $(\hbar\omega_\tau, \hbar\omega_t)$ = (1.584 eV, 1.584 eV) and (1.637 eV, 1.637 eV), respectively, and the estimated center position of the diagonal peak of component 2 [$(\hbar\omega_\tau, \hbar\omega_t)$ = (1.611 eV, 1.611 eV)], marked by purple crosses in Fig. 2a of the main text and in Supplementary Fig. 6.

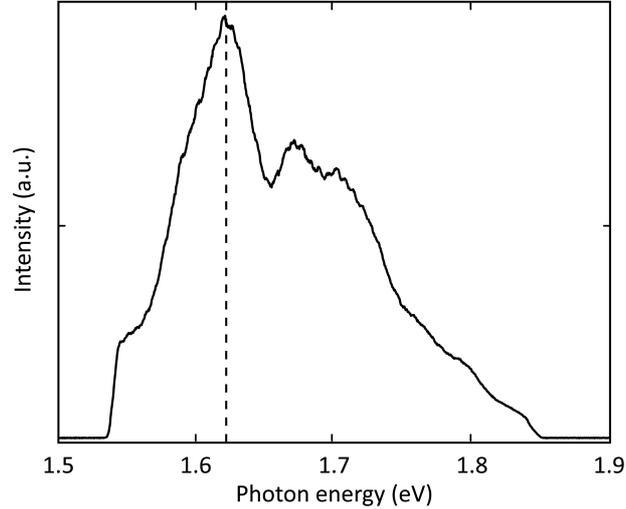

**Supplementary Figure 7:** Laser spectrum, with a structured peak with a maximum at 1.620 eV (dashed vertical line).

## Supplementary Note 3: Exclusion of sixth-order signal

We perform a 27-step (1×3×3×3) phase-cycling scheme, whereby not only the time delays between the four laser pulses, but also the phases of individual pulses are scanned. The PL signal is proportional to the final excited-state population (where we may exclude ESA pathways as explained in Supplementary Note 1) resulting from all possible Feynman pathways, which can be described as[1]:

$$p(\varphi_1, \varphi_2, \varphi_3, \varphi_4) = \sum_{\alpha,\beta,\gamma,\delta} \tilde{p}_{(n)}(\alpha,\beta,\gamma,\delta) \exp[i(\alpha\varphi_1 + \beta\varphi_2 + \gamma\varphi_3 + \delta\varphi_4)], \qquad (1)$$

where $\varphi_1, \varphi_2, \varphi_3, \varphi_4$, are the phases associated with the four excitation pulses and $\tilde{p}_{(n)}$ denotes the $n^{th}$-order contribution. The summation is carried out over integers $\alpha, \beta, \gamma, \delta$ within a range of $-\infty$ to $+\infty$ for each parameter, subject to the condition:

$$\alpha + \beta + \gamma + \delta = 0. \qquad (2)$$

A parameter combination [α, β, γ, δ] defines one specific nonlinear signal. E.g., [-1, 1, 1, -1] defines the fourth-order rephasing and [1, -1, 1, -1] the nonrephasing signal. To extract the signal with a specific contribution of [α, β, γ, δ] we take a discrete Fourier transform[1]:

$$\tilde{p}_{(n)}(\beta,\gamma,\delta) = \frac{1}{LMN}\sum_{n=0}^{N-1}\sum_{m=0}^{M-1}\sum_{l=0}^{L-1} p(l\cdot\Delta\varphi_{21}, m\cdot\Delta\varphi_{31}, n\cdot\Delta\varphi_{41})$$

$$\times \exp[-i(l\beta\cdot\Delta\varphi_{21} + m\gamma\cdot\Delta\varphi_{31} + n\delta\cdot\Delta\varphi_{41})], \quad (3)$$

where $L = 3$, $M = 3$, $N = 3$, are the numbers of steps we scan within a $2\pi$ range for the phase of each pulse, and $\Delta\varphi_{21}$, $\Delta\varphi_{31}$, $\Delta\varphi_{41}$ are the increments of the phase steps. In Supplementary Equation 3, $\alpha$ is missing because in the 1×3×3×3 phase-cycling scheme we fix $\varphi_1 = 0$, since the signal only depends on relative phase, i.e., $\varphi_{i1} = \varphi_i - \varphi_1$, for $i$ = 2, 3, 4.

In principle, 27-step phase cycling cannot exclude sixth-order signals. When extracting the desired fourth-order rephasing signal, we also obtain four types of sixth-order signals at the same time. They arise from combinations of [α, β, γ, δ] as[1]:

$$\alpha = +2, \beta = -2, \gamma = +1, \delta = -1, \quad (4)$$

$$\alpha = -1, \beta = -2, \gamma = +1, \delta = +2, \quad (5)$$

$$\alpha = +2, \beta = +1, \gamma = -2, \delta = -1, \quad (6)$$

$$\alpha = -1, \beta = +1, \gamma = -2, \delta = +2. \quad (7)$$

Similarly, when extracting the desired fourth-order nonrephasing signal, we also obtain four types of sixth-order signals at the same time. Their [α, β, γ, δ] combinations are[1]:

$$\alpha = -2, \beta = +2, \gamma = +1, \delta = -1, \quad (8)$$

$$\alpha = +1, \beta = +2, \gamma = -2, \delta = -1, \quad (9)$$

$$\alpha = -2, \beta = -1, \gamma = +1, \delta = +2, \quad (10)$$

$$\alpha = +1, \beta = -1, \gamma = -2, \delta = +2. \quad (11)$$

Although sixth-order signals normally are much weaker than fourth-order ones, as predicted by perturbative response function theory[12], they might still influence our analysis of the oscillating behavior of 2D maps. In order to examine any influence of the sixth-order signal overlapping with the fourth-order rephasing and nonrephasing signals, we conduct a separate measurement employing 64-step (1×4×4×4) phase cycling for $T$ = 50 fs. This allows us to separate the fourth-order rephasing and nonrephasing maps from eight sixth-order signals. The resulting absolute-valued 2D maps corresponding to these ten contributions are in Supplementary Fig. 8. The rephasing and nonrephasing fourth-order maps (left column) are consistent with those measured using 27-step phase cycling. There is no detectable sixth-order signal above the noise floor (right four columns) that would overlap with the fourth-order signals. Therefore, the oscillations as a function of $T$, discussed in the main text, arise from rephasing and nonrephasing fourth-order pathways without any higher-order signal.

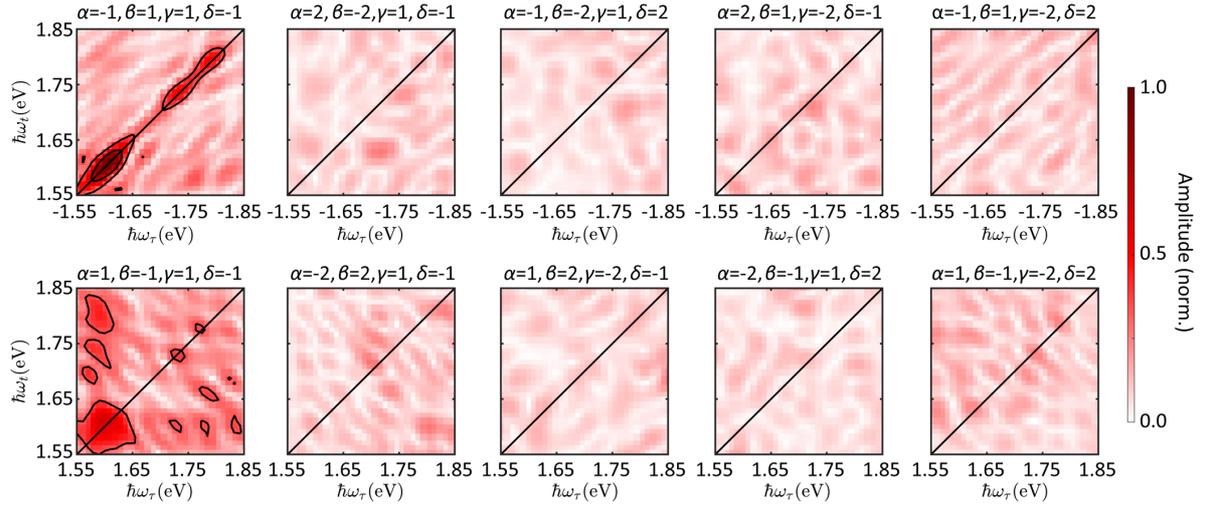

**Supplementary Figure 8:** Absolute-valued 2D maps of ten nonlinear contributions at *T* = 50 fs with 64-step phase cycling. [α, β, γ, δ] are labeled on top of the corresponding maps. The top left graph corresponds to the fourth-order rephasing signal and the bottom left one to the fourth-order nonrephasing one. There is no detectable signal beyond the noise floor for any of the other eight sixth-order nonlinear contributions.

## Supplementary Note 4: Reproducibility and noise-level analysis

The reproducibility of the 2D maps is confirmed by twice repeated 2DES measurements. The top two rows of Supplementary Fig. 9 show the rephasing 2D maps for *T* = 50, 500, 1000 fs measured on two different experimental runs (labeled A for the first and B for the second row) for ~20 μJ/cm$^2$. The 6 maps are all normalized to the maximum absolute value of the real part of the upper rephasing map for *T* = 50 fs. The evolution of the peak shape with *T* is consistent for the two measurements. The difference maps of A and B (third row of Supplementary Fig. 9) show only background noise, indicating that the contributing signals reproduce each other, thus cancelling each other in the difference.

      The reproducibility can be better illustrated by comparing the amplitude evolutions of an exemplary pixel (marked by the green diamond in the upper left panel of Fig. 2a of the main text) as a function of *T* for the two measurements (Supplementary Fig. 10). The green curve (the first run) and the blue curve (the second run) agree with each other.

      The difference map between two measurements for the same *T* provides a way to evaluate the noise level, by evaluating the standard deviation (SD) of the data. However, this inherits the noise from both maps, hence we need to estimate correctly the signal-to-noise ratio of each individual map.

      Supplementary Figs. 11a,b are zoomed-out rephasing maps for *T* = 50 fs, corresponding to the data sets in the first column of Supplementary Fig. 9. We analyze separately the noise level of each map by evaluating SD outside the signal region marked by the dashed orange box, yielding 0.0745 and 0.0788 for panels a and b, respectively.

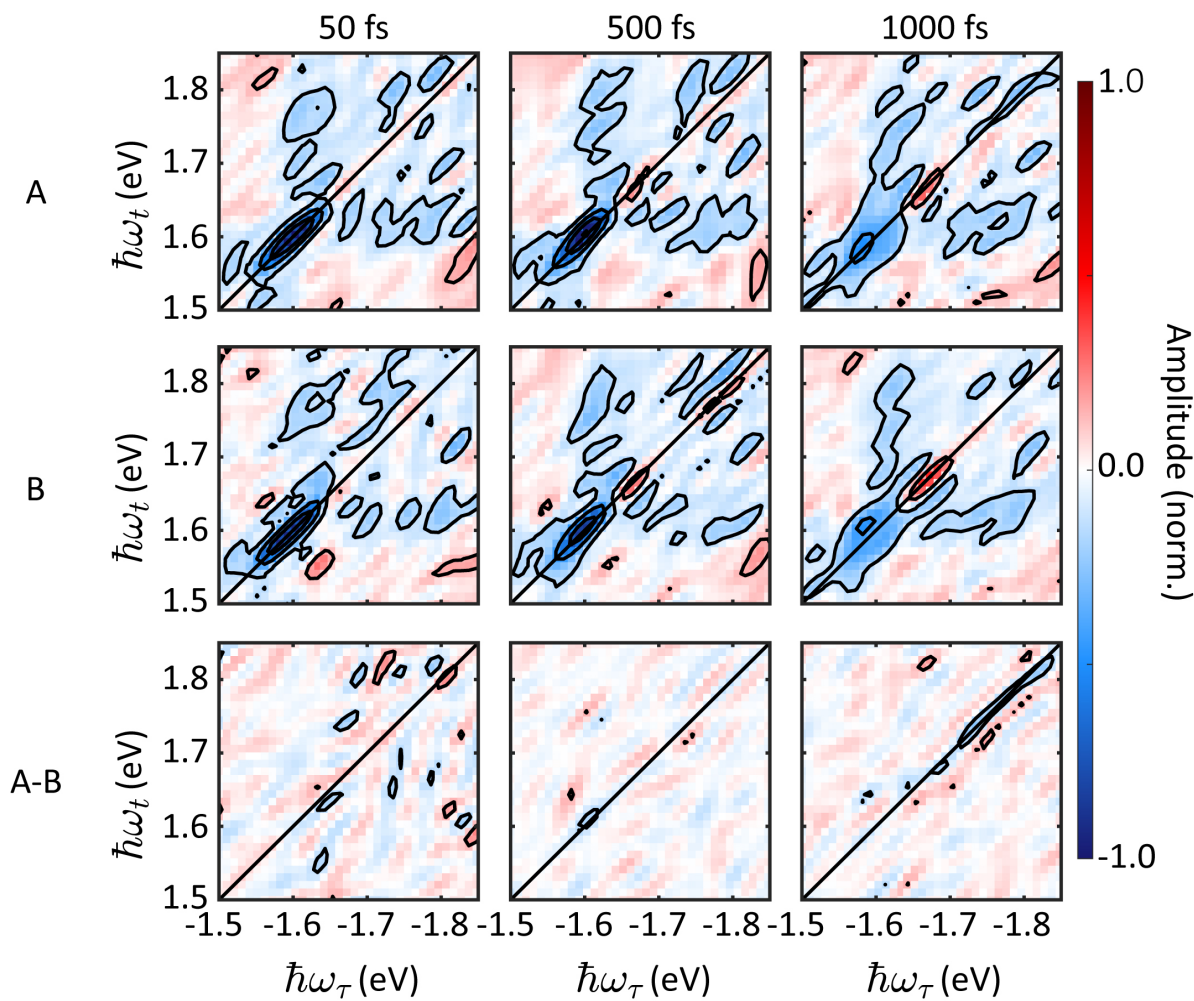

**Supplementary Figure 9:** Reproducibility of rephasing 2D maps at $T$ = 50, 500, 1000 fs. The row labeled A contains a first and row B a second set of measurements. The evolution of peak shape with respect to $T$ is consistent for the two measurements. Row A-B contains the difference between A and B. The signals largely cancel, leaving only background noise.

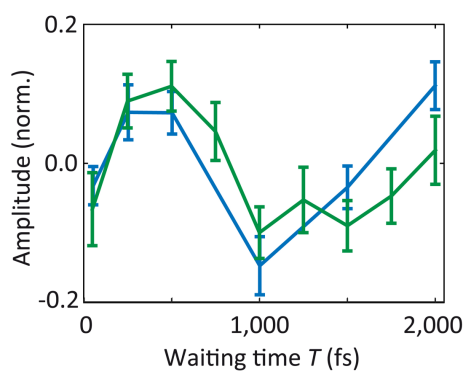

**Supplementary Figure 10:** Reproducibility of time traces for the same position within the rephasing 2D maps measured in two repeated runs. The error bars are evaluated by calculating the fluctuations within a region containing background noise.

The difference map is in Supplementary Fig. 11c. SD inside the dashed orange box of the difference map is 0.0686, close to the calculated SD from panels a, b outside of the orange box. The agreement between SD from the outside-box region in panels a, b and from the inside-box region in panel c indicates that the noise is evenly distributed. Thus, the SD calculated outside the signal region can be used to evaluate the fluctuation ranges of the measured amplitude for each single pixel in the signal region. For each *T*, we separately extract SD, and use it to create an error bar for the corresponding *T* step in the amplitude evolution curves in Supplementary Figs. 10,12 and Fig. 2c of the main text.

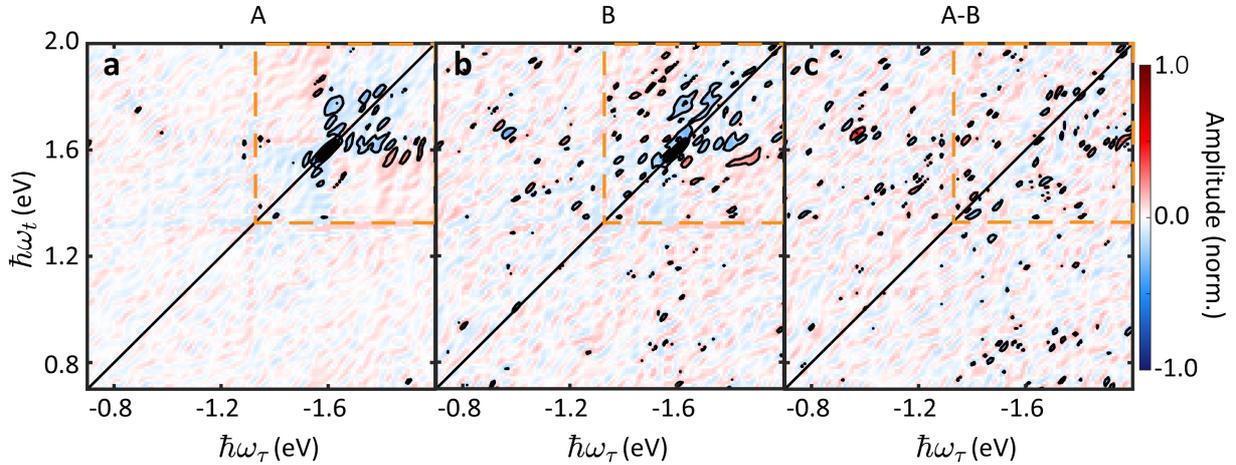

**Supplementary Figure 11:** Noise-level analysis. **a,b**, Zoomed-out rephasing 2D maps at *T* = 50 fs of the first column of Supplementary Fig. 9 for (a) row A and (b) row B. The dashed orange boxes mark the signal regions in the two maps. **c**, Difference between panels a and b.

## Supplementary Note 5: Exclusion of biexciton signal

Biexcitons have a binding energy of ~20 meV[13–15]. In view of the Feynman-pathway analysis of Supplementary Figure 2, biexcitons can only be detected through excited-state absorption (ESA) pathways in a 2D spectroscopy measurement, meaning that they appear in 2D maps as a peak outside the diagonal and at lower probe energies (~20 meV) compared to the neutral exciton[13]. This should correspond to a strongly asymmetric lineshape towards the red (low detection frequency $\omega_t$) in the absorptive 2D maps. We do not observe this, indicating that the effect of biexcitons is negligible in our experiments at room temperature.

The following two factors could explain this: 1) The binding energy of the neutral biexciton is an order of magnitude lower than the exciton binding energy[16,17]. Thermal fluctuations make biexcitons unstable, and lead to biexciton dissociation at room temperature. Therefore, upon laser excitation, even if some biexcitons are present at room temperature, we expect their spectral signal to be much weaker than the single neutral exciton. 2) As discussed in Supplementary Note 1, in population-detected 2D spectroscopy, the two types of excited-state absorption (ESA) pathways, i.e., ESA 1 and ESA 2, usually cancel each other to some extent (depending on their associated fluorescence quantum yields), leading to a reduction of ESA signal, hence a further reduction of the contribution from biexcitons.

# Supplementary Note 6: Extracting the oscillation period

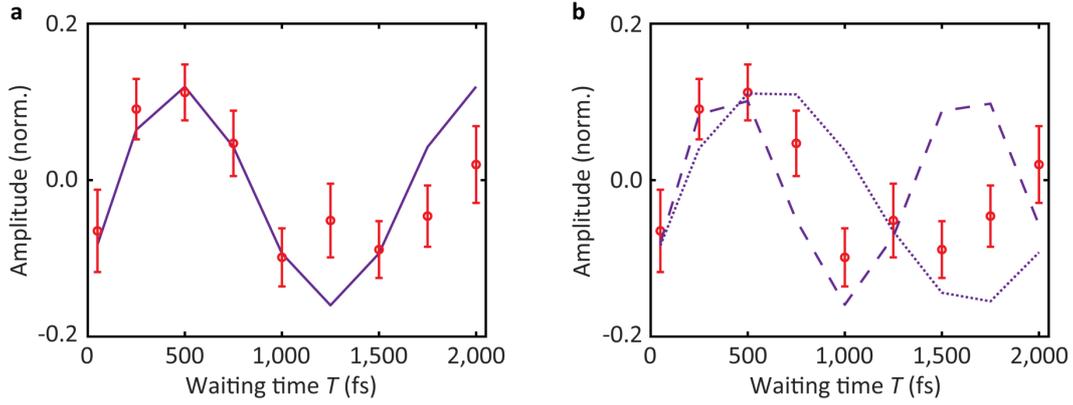

**Supplementary Figure 12: a**, Fitted oscillating curve (solid purple). **b**, Exemplary sinusoidal curves with varied oscillation periods of 133 fs (dotted) and 139 fs (dashed) yielding significant deviations with respect to the measurements, thus demonstrating the accuracy of the fitting of panel a. The error bars are evaluated by calculating the fluctuations within a region containing background noise (Supplementary Note 4).

To extract the oscillation period from the amplitude evolution curve (shown in Fig. 2c in the main text), we fit it using a sine function:

$$Y = A \sin(2\pi(x - x_0)/w),  \tag{12}$$

with amplitude $A$, phase $x_0$, and period $w$ restricted to ~118–230 fs (as obtained from the positions of the constituent components). An oscillation period $w\sim 136$ fs is obtained from the fitting. Supplementary Fig. 12a shows the fitting results, where the measured data are plotted as red circles with error bars. The purple curve is obtained by taking only those time points (50, 250, 500, 750, 1000, 1250, 1500, 1750, 2000 fs) at which 2D maps are measured, and connecting the values obtained from the fitting function by straight line segments. The accuracy of the extracted period is checked by varying it and comparing the resulting curves with the data. As shown in Supplementary Fig. 12b, if we change the period to either 133 fs (dotted purple line) or 139 fs (dashed purple line), the curves deviate strongly from the data, from which we derive the ±2 fs error of the main text.

# Supplementary Note 7: Definition of Huang–Rhys factor *S*

We consider an electronic (or excitonic) ground state |g⟩ and an electronic (or excitonic) first excited state |e⟩. Using a harmonic oscillator to approximate the dependence of potential energy on a vibrational (phonon) dimensionless coordinate $q$ with the ground-state minimum at $q = 0$ [12],

$$V_g(q) = \frac{\hbar\omega}{2} q^2, \tag{13}$$

the potential curvature leads to a vibrational level spacing of $\hbar\omega$, where $\omega$ is the phonon angular frequency, creating sublevels $|g_i\rangle$, i = 0, 1, 2, …. For full information on the system, we also need to describe the excited-state potential[12],

$$V_e(q) = \hbar\omega_{eg} + \frac{\hbar\omega}{2}(q + d)^2, \tag{14}$$

in which we assume, for simplicity, the same curvature, thus the same $\hbar\omega$ as in the ground state, a (vertical) energy difference $\hbar\omega_{eg}$, and a (horizontal) shift along the phonon coordinate that can be formulated in a dimensionless way, *d*, between the two potential minima. This generates sublevels $|e_i\rangle$, i = 0, 1, 2, …. Substituting Supplementary Equation 13 into Supplementary Equation 14 leads to

$$V_e = \hbar\omega_{eg} + V_g + \hbar\omega S + \hbar\omega q d, \tag{15}$$

where $S = d^2/2$ is the Huang–Rhys factor, and $\hbar\omega S$ is the reorganization energy.

## Supplementary Note 8: Peak position of individual Feynman pathways

The peak position along the (horizontally displayed) $\hbar\omega_\tau$ excitation energy axis can be found by evaluating the energy difference between the states of the coherence created after the first light-field interaction. E.g., considering pathway 1, the first coherence created (and evolving with $\tau$) is $|g_0\rangle\langle e_0|$. According to Fig. 3a of the main text (black arrow), this is located at the intermediate of the three possible transition energies for the excitation, i.e., at $\hbar\omega_\tau$, where the minus sign in Fig. 3e of the main text arises from the definition of sign of the frequency of a coherent state (a coherent state $|X\rangle\langle Y|$ has positive frequency when level $|X\rangle$ is higher in energy than level $|Y\rangle$ and negative frequency if level $|X\rangle$ is lower than $|Y\rangle$, see Supplementary Note 1). Likewise, the coherence after the third interaction (and evolving with *t*) is $|e_0\rangle\langle g_1|$, at the lowest of the 3 transition energies $\hbar\omega_1$ for the detection, according to the blue arrow in Fig. 3a of the main text with positive sign because $|e_0\rangle$ is higher than $|g_1\rangle$. All other peaks are assigned in a similar way, so that the displayed pattern emerges.

## Supplementary Note 9: 2D beating maps for different *S* values

Supplementary Figure 13 contains simulated 2D beating maps for various *S*, a subset of which is shown in Fig. 4a of the main paper. The lowest contour lines of the experimental and the simulated beating maps in Figs. 4a,b of the main text and Supplementary Figure 13 show some "jagged" behaviour. There are several factors that could contribute to this. 1) The measurement uncertainty arising from noise becomes more visible at the lowest contour line for any given signal-to-noise level, leading to deviations from an ideal elliptical shape. 2) The energy resolution is given by the temporal scanning range and is ~40.6 meV. Using additional four-fold zero padding, one pixel has a side length ~10 meV, corresponding to 20 frequency pixels in the spectral ranges displayed in Fig. 4 of the main text, along either frequency axis. Thus, any (random) deviation, due to noise, in just one or two neighbouring independent frequency intervals will lead to a "jagged" outline of the respective contour line because there are only few points that make up any such line. 3) The beating maps represent cuts through a three-dimensional Fourier space for a particular $\omega_T$. However, the experimental scanning procedure sets a finite resolution along the $\omega_T$ direction, and any beating contribution has a finite width along this axis. Thus, it is possible that contributions from several different beating frequencies overlap at any given $\omega_T$. If different beating contributions are located at different ($\omega_\tau$, $\omega_t$) positions, their interference can lead to a more complex appearance of the beating map for any particular $\omega_T$ cut position. 4) The spectra are influenced by the shape of the excitation laser spectrum. If this spectrum deviates from a perfectly smooth function (such as a Gaussian), this will introduce additional structure. For optimal comparison between theory and experiment, we use the experimental spectrum also for simulations, thus jagged contour lines can emerge even in simulations without noise.

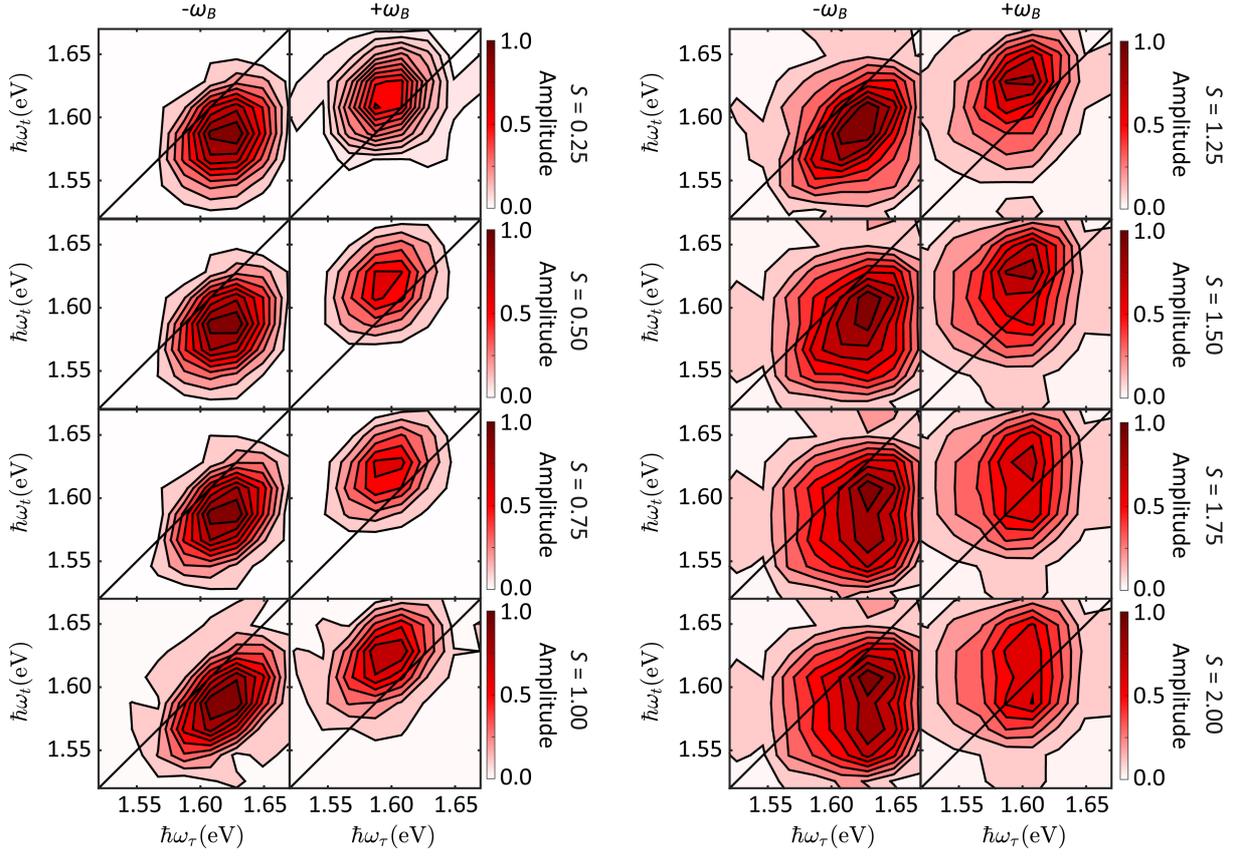

**Supplementary Figure 13:** Simulated 2D beating maps for $-\omega_B$ (left side in each double column) and $+\omega_B$ (right side in each double column) and for different $S$ values.

## Supplementary Note 10: Sample temperature during 2DES measurement

Heating the sample through laser irradiation during the experiment may cause thermal instabilities or damage. In addition, sample temperature is a decisive factor for calculating the 2D beating maps, because it determines the distribution of the initial population of the ground and excited vibrational states. We estimate the sample temperature during the 2DES measurements by adapting the two-temperature model[18], for the coupling of electronic and vibrational degrees of freedom in solids. This describes the energy transfer inside a material with two coupled generalized heat conduction equations for the temperature of the electrons $T_e$ and the lattice $T_l$,

$$c_e \frac{\partial T_e(r,t)}{\partial t} = k_e \frac{\partial^2 T_e(r,t)}{\partial r^2} - \alpha\big(T_e(r,t) - T_l(r,t)\big) + \sigma \cdot I(r,t), \qquad (16)$$

$$c_l \frac{\partial T_l(r,t)}{\partial t} = k_l \frac{\partial^2 T_l(r,t)}{\partial r^2} + \alpha\big(T_e(r,t) - T_l(r,t)\big) - \beta(T_l(r,t) - T_s), \qquad (17)$$

where $r$ is a spatial lateral coordinate, $t$ the time, $c_e$ and $c_l$ are the electron and lattice volumetric heat capacities, $k_e$ and $k_l$ are the electron and lattice thermal conductivities, $\alpha = c_e/\tau_r$ is the thermal coupling function between electron and lattice subsystems, and $\tau_r$ the characteristic time of electron gas cooling due to energy exchange with the lattice[18]. $\beta$ characterizes the rate of energy exchange between 1L-MoSe$_2$ and substrate. This can be expressed as the product of the interfacial thermal conductance $h_c$ between 1L-MoSe$_2$ and substrate and the laser-irradiated area $A$. $I(r,t)$ is the intensity of the laser beam, $\sigma$ is the absorbance of the sample. We set $T_s \equiv 300$ K, by assuming infinitely fast heat dissipation from the substrate to surrounding areas.

To simplify the calculation, we ignore transverse thermal diffusion by setting $k_e = k_l = 0$, so that the absorbed laser energy is confined in the irradiated volume $V = A \cdot l$ before it is transferred to the substrate ($l$ is the thickness of 1L-MoSe$_2$), and Supplementary Equations (16) and (17) reduce to

$$c_e \cdot V \cdot \frac{\partial T_e(t)}{\partial t} = -(c_e/\tau_r) \cdot V \cdot \big(T_e(t) - T_i(t)\big) + \sigma \cdot I(t), \tag{18}$$

$$c_i \cdot V \cdot \frac{\partial T_i(t)}{\partial t} = (c_e/\tau_r) \cdot V \cdot \big(T_e(t) - T_i(t)\big) - h_c \cdot A \cdot (T_i(t) - T_s). \tag{19}$$

To take into account the cumulative effects owing to the high repetition frequency (80 MHz) of the laser, we consider a multi-pulse heating model[19]:

$$I(t) = \sum_{n=0}^{+\infty} I_0 \cdot e^{-\frac{(t - n \cdot t_i)^2}{w^2}} \tag{20}$$

for the laser intensity irradiating the material, with $I_0$ representing the peak intensity, $w$ the duration of every laser pulse. The time interval $t_i$ = 12.5 ns between individual pulses is defined by the repetition frequency. The integer number $n$ ranges from zero to infinity, so that the sample is continuously heated from pulse to pulse. Because the laser power of the 2DES measurement is much lower than the damage threshold of the material, we calculate $c_e$ using[18]:

$$c_e = \gamma \cdot T_e, \tag{21}$$

where $\gamma$ is a proportionality constant that connects the heat capacity of the electron gas with its temperature. The values for all parameters in Supplementary Equations 18–21 are in Supplementary Table 1.

Supplementary Fig. 14a plots the evolution of $T_e$ (red curve) and $T_l$ (blue curve) within one interval (12.5 ns) between two laser pulses. Upon the arrival of the first pulse, $T_e$ starts to increase since electrons are excited. Then the energy is transferred from electrons to lattice, resulting in a subsequent rise of $T_l$. The lattice finally gives its energy to the substrate because of the thermal contact between them, hence $T_l$ decreases. If the interfacial thermal conductance $h_c$ is high enough, $T_l$ will drop back to the initial temperature before the next pulse comes, thus the same circle will repeat between any two pulses. On the other hand, if the heat released to the lattice does not have time to fully dissipate to the substrate before the next pulse arrives, the cumulative effects will lead to an increase of $T_l$ from pulse to pulse until an equilibrium value is reached, as shown in Supplementary Fig. 14b, that plots $T_l$ at the arrival time of pulse number $n$ ($n = 1, 2, 3, ...$). Calculation indicate that for ~3.6×10$^{-14}$ J used in our 2DES measurement, $T_l$ increases from 300 to ~308 K within the first 100 ns, then remains constant. Thus, there is no unwanted heating of the sample, thermal instabilities or damage.

There are two key assumptions in our calculation. 1) We assume that the absorbed energy is confined within the region of the laser focus, and does not diffuse to surrounding areas. 2) From a 1–2 orders of magnitude disagreement in the literature on $h_c$ of TMDs (ranging between 0.1 to 14 MW m$^{-2}$ K$^{-1}$ [20–22]), we use the minimum in our calculation, so that the maximum possible $T_l$ can be calculated by our model. Because both assumptions overestimate the equilibrium $T_l$, these assumptions ensure that the experimental $T_l$ does not exceed the calculated ~308 K, with a negligible heating during our measurements.

**Supplementary Table 1:** Parameters used in the TTM calculations.

| Parameter | Value | Parameter | Value |
|---|---|---|---|
| $c_l$ | $1.87 \times 10^6$ Jm$^{-3}$K$^{-1}$ [23] | $\sigma$ | 2.5% [24] |
| $A$ | $5.3 \times 10^{-14}$ m$^2$ | $I_0$ | 2.8 W |
| $l$ | 0.65 nm | $T_r$ | 12.5 ns |
| $\tau_r$ | $240 \times 10^{-15}$ s [18] | $t_c$ | 100 fs |
| $h_c$ | 0.1 MWm$^{-2}$K$^{-1}$ [21] | $w$ | 7.2 fs |
| $\gamma$ | 67.6 J m$^{-3}$ K$^{-2}$ [18] | | |

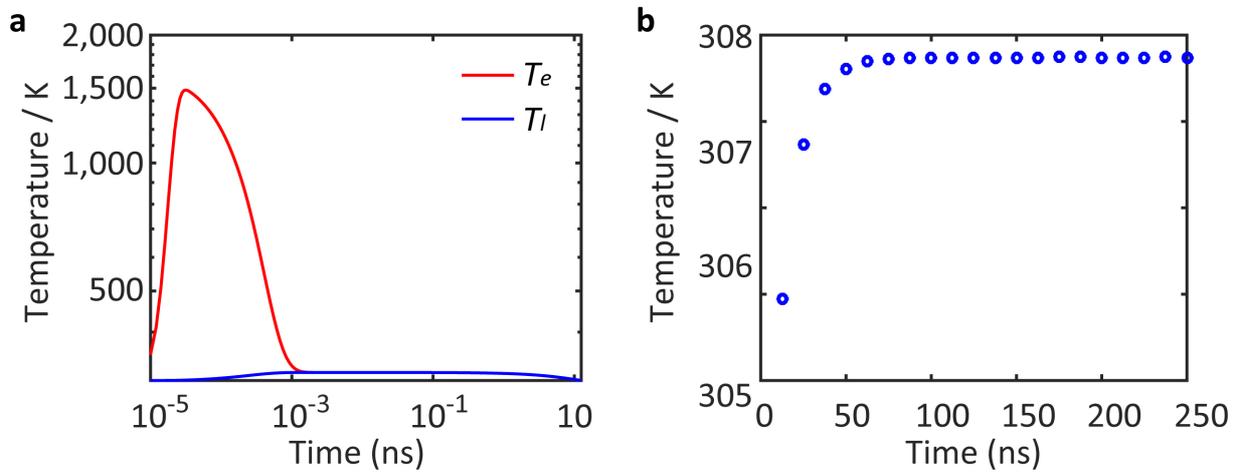

**Supplementary Figure 14: a,** Evolution of the temperature of the electrons $T_e$ (red curve) and the temperature of the lattice $T_l$ (blue curve) within one interval (12.5 ns) between two laser pulses. **b,** Calculated $T_l$ at the arrival time of pulse $n$ ($n$ = 1, 2, 3, …).

## Supplementary Note 11: Autocorrelation measured at the sample position

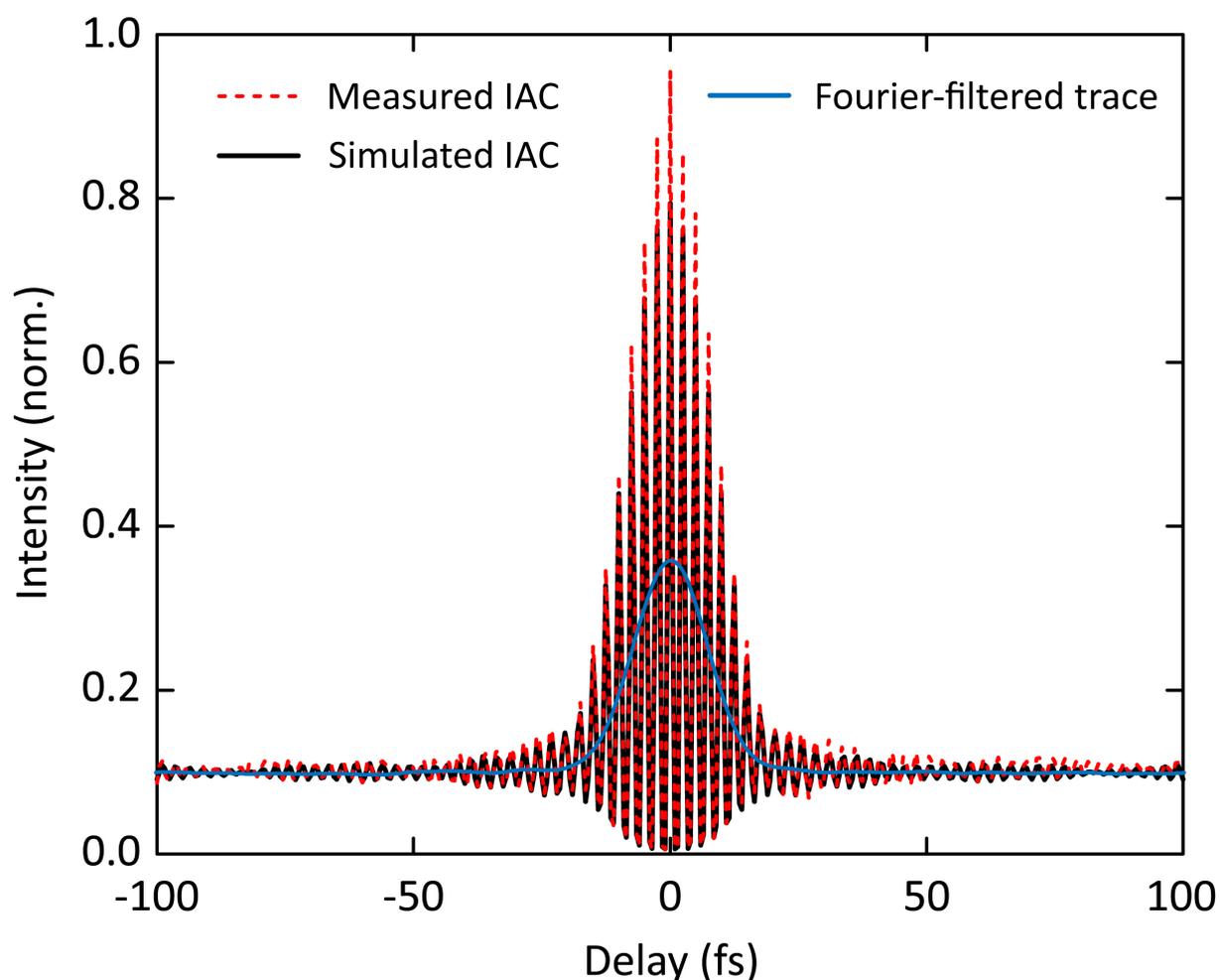

**Supplementary Figure 15:** Measured (dashed red) and simulated (solid black) interferometric autocorrelation (IAC) assuming a flat spectral phase and the separately measured laser spectrum. The pulse duration of ~12 fs can be obtained by dividing the full width at half maximum (FWHM) of the Fourier-filtered trace by $\sqrt{2}$ (solid blue curve)[3].

## Supplementary References

1. Tan, H.-S. Theory and phase-cycling scheme selection principles of collinear phase coherent multi-dimensional optical spectroscopy. *J. Chem. Phys.* **129**, 124501 (2008).

2. Draeger, S., Roeding, S. & Brixner, T. Rapid-scan coherent 2D fluorescence spectroscopy. *Opt. Express* **25**, 3259–3267 (2017).

3. Goetz, S., Li, D., Kolb, V., Pflaum, J. & Brixner, T. Coherent two-dimensional fluorescence micro-spectroscopy. *Opt. Express* **26**, 3915–3925 (2018).